\documentclass{article}

\usepackage{amsfonts}
\usepackage{amssymb}
\usepackage{amsmath}
\usepackage{pgfplots}
\usepackage{tikz}
\usepackage{booktabs}
\usepackage{algorithm}
\usepackage[noend]{algpseudocode}
\usetikzlibrary{arrows,automata}
\usepackage{hyperref}
\usepackage{float}
\usepackage{verbatim}
\usepackage{graphicx}
\usepackage{caption}
\usepackage{subcaption}
\usepackage{epstopdf}
\usepackage{epsfig}
\usepackage{url}
\usepackage{array}
\usepackage{lipsum}
\usepackage{tabularx}
\usepackage{multirow}
\usepackage{authblk}
  \colorlet{anglecolor}{green!50!black}
  \colorlet{sincolor}{red}
  \colorlet{tancolor}{orange!80!black}
  \colorlet{coscolor}{blue}
\tikzstyle{decision} = [diamond, draw, fill=blue!20, 
    text width=4.5em, text badly centered, node distance=3cm, inner sep=0pt]
\tikzstyle{block} = [rectangle, draw, fill=blue!20, 
    text width=5em, text centered, rounded corners, minimum height=4em]
\tikzstyle{line}=[draw, line width=1.5pt]
\tikzstyle{arrow}=[draw, -latex, line width=1.5pt] 
\tikzstyle{cloud} = [draw, ellipse,fill=red!20, node distance=3cm,
    minimum height=2em]
  \tikzstyle{important line}=[very thick]
  \tikzstyle{information text}=[rounded corners,fill=red!10,inner sep=1ex]

\output\expandafter{\the\output\floatfix}
\newcolumntype{P}[1]{>{\centering\arraybackslash}p{#1}}
\makeatletter
\def\floatfix{%
\expandafter\ifx\csname r@x@one\endcsname\relax
\else
\ifnum\c@page=\numexpr\expandafter\expandafter\expandafter
              \@secondoftwo\csname r@x@one\endcsname-1\relax
\aftergroup\figone
\fi
\fi}
\makeatother

\makeatletter
\def\Demo{%
  \def\Ginclude@graphics##1{%
      \rule{\@ifundefined{Gin@@ewidth}{150pt}{\Gin@@ewidth}}%
      {\@ifundefined{Gin@@eheight}{100pt}{\Gin@@eheight}}}}
\makeatother
\makeatletter
\def\@copyrightspace{\relax}
\makeatother

\newcommand{\svm}{\textbf{SVM} }
\newcommand{\misvm}{\textbf{MI-SVM} }
\newcommand{\rmilNOR}{\textbf{rMIL\textsuperscript{nor}} }
\newcommand{\rmilAvg}{\textbf{rMIL\textsuperscript{avg}} }
\newcommand{\gicf}{\textbf{GICF} }
\newcommand{\nmil}{\textbf{nMIL} }
\newcommand{\nmildelta}{\textbf{nMIL\textsuperscript{$\Delta$}}}
\newcommand{\nmilseq}{\textbf{nMIL\textsuperscript{$\Omega$}}}

\newcommand{\super}{r}
\newcommand{\bag}{i}
\newcommand{\doc}{j}
\newcommand{\lead}{k}

\newcommand{\evt}{$e\,$}
\newcommand{\emp}[1]{{ \textbf{#1}}}
\title{Modeling Precursors for Event Forecasting\\
via Nested Multi-Instance Learning} 

\author{
  Yue Ning\textsuperscript{1},
  Sathappan Muthiah\textsuperscript{1},
  Huzefa Rangwala\textsuperscript{2},
  Naren Ramakrishnan\textsuperscript{1}
}
\affil{\textsuperscript{1}Discovery Analytics Center, Virginia Tech, Arlington, VA 22203\\
\textsuperscript{2}Department of Computer Science, George Mason University, Fairfax, VA 22030}

\date{}
\begin{document}
\maketitle

\begin{abstract}
Forecasting events like civil unrest movements, disease outbreaks, 
financial market movements and government elections 
from open source indicators such as news feeds and 
social media streams is an important and challenging problem. 
From the perspective of human analysts and policy makers, 
forecasting  algorithms need to provide supporting evidence and 
identify the causes related to the event of interest. We develop a novel 
multiple instance learning based approach that jointly 
tackles the problem of 
identifying evidence-based precursors and forecasts events into the future. 
Specifically, given a collection of streaming 
news articles from multiple sources we develop a nested multiple instance learning 
approach to forecast significant societal events across three countries in Latin America. 
Our algorithm is able to identify news articles considered as 
precursors for a protest. Our empirical evaluation 
shows the strengths of our proposed approaches in filtering candidate 
precursors, forecasting the occurrence of events with a lead time and 
predicting the characteristics of different events in comparison 
to several other formulations. 
We demonstrate through case studies the effectiveness of our proposed 
model in filtering the candidate precursors for inspection by a human analyst. 
\end{abstract}

\section{Introduction}

Forecasting societal uprisings, civil unrest movements, and terror threats 
is an important and challenging problem. Open source data sources 
(e.g., social media and news feeds) have been known to serve as surrogates
in forecasting a broad class of events, e.g.,
disease outbreaks \cite{Arhrekar:2011}, election outcomes \cite{OConnorBRS:10,tumasjan2010predicting}, stock 
  market movements \cite{Bollen2011}  and protests \cite{Ramakrishnan:2014}.
While many of these works focus on predictive performance, there is
a critical need to develop methods that also yield insight by
identifying precursors to events of interest. 

This paper focuses on the problem of identifying precursors (evidence)  for forecasting 
significant societal events. Modeling and identifying the precursors for a given protest is useful 
information for the human analyst and  policy makers  as it discerns the underlying reasons behind 
a civil unrest movement. 
Specifically, the objective of this study is to forecast protest across different cities in three Latin American 
countries (Argentina, Brazil and Mexico). 6000 news outlets are tracked daily across these countries with the goal 
of forecasting the occurrence of a protest with atleast one day of lead time. From the news feeds, we also aim to 
identify the specific news articles that can be considered as precursors for the targer event. 

We formulate the precursor identification and forecasting problem within a novel multiple instance learning algorithm (MIL). 
Multiple instance learning algorithms~\cite{andrews2002support, zhou07} are a class of supervised learning techniques that have labels for a group of instances, but 
not for individual instances. We make a similar mapping for the group of collected news articles to have no individual class label associated with
every news article but the group of news articles are attached with a label indicating the occurrence of a protest. We extend the standard 
MIL algorithm by introducing a nested structure, where we group news articles published in a given day at the first  level and 
then group the collection of individual days at the second level. 
This nested MIL approach allows for  modeling the sequential constraints between the  news articles (grouped by days) 
published on different days and also provides a probabilistic estimate for every news article and the collection of news article. This estimate 
is signficant because it indicates for a given news article the probability of it signaling a protest event.  Note, in our datasets we do not have any 
training labels to indicate the protest indicator per news article.

\begin{figure*}[h]
\includegraphics[width=1.0\linewidth]{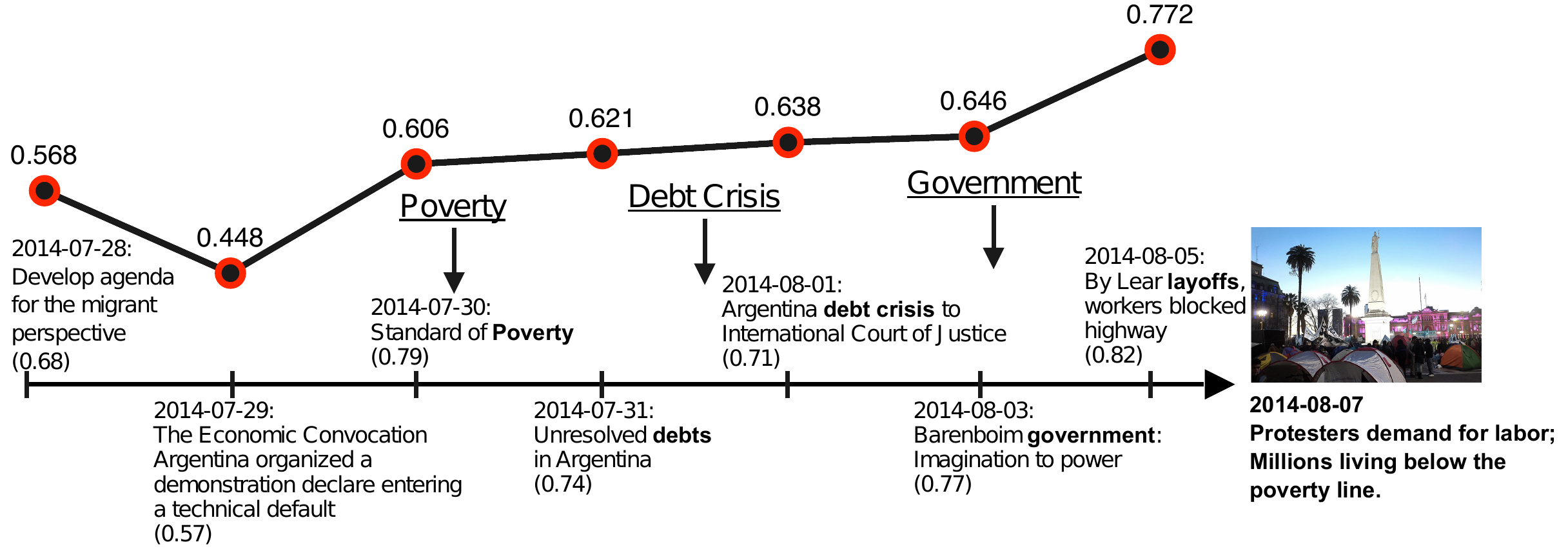} 
\caption{Precursors story line for a protest event in Argentina. The x-axis is the timeline. The dots above with numbers are the probabilities for each day that the model generated for the target event. Each precursor document is titled in the timeline. }\label{fig:precursor_demo}
\end{figure*}

Figure~\ref{fig:precursor_demo} shows an example of precursors detected by our model. 
On the right of the timeline, is a news report about a protest event in Argentina.
The connected dots denotes the generated probability of this day being positive.
From this example, we find that within 10 days
before the event, there are multiple similar reported events that were 
selected as
highly probable leading indicators or evidence. Most of 
planned societal events are a consequence of several factors that affect the
different entities within communities and their relationships with each other (or the 
government) over time. In this specific example, the leading precursor some days before 
the protest was an article commenting on standards of living in Argentina and rising poverty levels. 
The International Court of Justice also provided a verdict on the debt crisis. All these factors 
led to the final protest where general population demanding work opportunities.


The main contributions of this study are summarized as follows:
\begin{enumerate}
  \item \textbf{A novel
    nested framework of multi-instance learning for event
    forecasting and precursor mining.}
    We formulate event forecasting and precursor mining for multiple cities in a country as a
    multi-instance learning problem with a nested structure. By
    estimating a prediction score for each instance in the history data,
    we automatically detect significant precursors for different events.
  \item \textbf{Harness temporal constraints in multi-instance learning.}
    We explore different penalty function and regularizations where we employ the
    temporal information in our   dataset under
    assumption that most events of interest are follow-up reports of
    other events that happened before, and most planned events are
    developing over time. 
\item \textbf{Modeling for various event categories in multiple geo-locations}. 
We extend the nested MIL formulation for general purpose multi-class classification to determine 
necessary attributes of events in terms of their underlying population. 
  \item \textbf{Application and evaluation with comprehensive experiments.}
    We evaluate the proposed methods using  news data collected from
    July 2012 to December 2014 in three countries of Latin America:
    Argentina, Mexico, and Brazil. For comparison, we implement other
    multi-instance algorithms, and validate the effectiveness and efficiency
of the proposed approach. We also perform qualitative and quantitative 
    analysis on the precursors inferred by our model.
\end{enumerate}


The rest of this paper is organized as follows. We discuss related work in Section~\ref{sec:related}.
Section~\ref{sec:problem} introduces the problem setup and our proposed model based on
Multi-Instance Learning is presented in Section~\ref{sec:method}. This section is then followed by experiments
and evaluations on real world datasets presented in
Section~\ref{sec:experiment} and Section~\ref{sec:result} . 
Finally, we concludes with a summary of the
research in Section~\ref{sec:conclusion}.


\section{Related Work}\label{sec:related}
 
\textbf{Event Detection and Forecasting.}
Event detection and forecasting from online open source datasets has been
an active area of research in the past decade. 
Both supervised and unsupervised machine learning techniques 
have been developed to tackle different challenges. 
Linear regression models 
use simple features  to predict the occurrence time of 
future events~\cite{Arias:2014, Bollen2011, He:2013, OConnorBRS:10}.
Advanced techniques use a combination of sophisticated features 
such as topic related keywords, as input to support vector machines, 
LASSO and multi-task learning approaches ~\cite{Wang:2012, Ritterman09}.
Ramakrishnan \textit{et al.}~\cite{Ramakrishnan:2014} designed 
a framework (EMBERS) 
for predicting civil unrest events in different locations by using a wide combination of 
models 
with heterogeneous input sources ranging from social media to satellite images. 
Zhao \textit{et al.}~\cite{Zhao:2015} combine multi-task learning and dynamic features from social
networks for spatial-temporal event forecasting. Generative models have 
also been used in~\cite{zhao15:sdm} to jointly model 
the temporal evolution in semantics and geographical burstiness
within social media content. Laxman \textit{et al.}~\cite{Laxman3:2008} designed a 
generative model for categorical event prediction in event streams using frequent episodes. 
However, few existing approaches provide evidence 
and interpretive analysis as support for event forecasting.

\textbf{Identifying Precursors.}
Identifying precursors for significant events is 
an interesting topic and has been used extensively 
for interpretive narrative generation and in storytelling algorithms \cite{Hossain:2012}.
Rong \textit{et al.}~\cite{Rong:2015:WHI} developed a  combinational mixed Poisson process (CMPP) model to
 learn social, external and intrinsic influence in social networks.

\textbf{Multiple Instance Learning.}
In the multiple instance learning (MIL)  paradigm, we are given 
labels for sets of instances commonly referred as \emph{bags} or \emph{groups}. However, individual instance-level 
labels are unknown or missing. The bag-level labels are assumed to be an association function 
(e.g., OR, average) of the 
unknown instance level labels. One approach to MIL adapts support vector machines (SVMs) by:
(i) modifying the maximum margin formulation to 
 discriminate between bags rather than individual instances \cite{andrews2002support}, and 
(ii) developing kernel 
 functions that operate directly on bags \cite{Gartner2002}. 
Other multiple instance learning approaches 
and various applications are found in a detailed survey \cite{amores2013multiple}.
Specifically, the generalized MIL \cite{Weidmann2003mil} formulation assumes the presence of 
multiple concepts and  a
bag is classified as positive if there exists instances from
every concept. 
Relevant to our work, besides predicting bag labels, Liu \textit{et al.} \cite{jmlr-LiuWZ12} seek 
to identify the  key  instances within 
the positively-labeled bags using nearest neighbor techniques. 
Recent work \cite{Kotzias:2015:KDD} has
focused on instance-level predictions from group labels (GICF) and 
allowed for the application of general aggregation functions with applications
to detecting sentiments
associated with sentences within reviews. 

The methods proposed in this paper can be viewed as complementary to prior work, casting the 
forecasting and precursor discovery problems within novel 
extensions of multiple instance learning.


%

\section{Problem Formulation} \label{sec:problem}


\begin{table}
\small
  \centering
  \caption{Notations}\label{tab:notations}
  \begin{tabular}{lm{10cm} }
   \toprule
   \hline
   \textbf{Variable} & \textbf{Meaning} \\
  \midrule
  $\mathcal{S} = \{\mathbb{S}\} $& a set of $n$ super bags in our dataset\\
  \hline
$\mathbb{S} =[\mathcal{X}_{\bag}], \bag \in \{1,...,t\}$ & an ordered set of $t$ ``bags'' in $\mathbb{S}$\\
  \hline
$\mathcal{X}_{\bag} = \{\mathbf{x}_{\bag\doc}\}, \doc \in \{1,...,n_{\bag}\}$& a set of instances with $n_{\bag}$=$|\mathcal{X}_{\bag}|$, number of instance in a bag $\mathcal{X}_{\bag}$\\
\hline
$\mathbf{x}_{ij} \in \Re^{V\times 1}$ & the $\doc$-th instance in set $\mathcal{X}_{\bag}$, a V-dimension vector\\
\hline
$Y \in \{-1, +1\}$ & label of super bag\\
\hline
$P \in [ 0, 1]$& estimated probability for a super bag\\
\hline
$\mathbb{P}_{\bag} \in [0, 1]$ & the probability of bag $\bag$ in super bag to be positive\\
\hline
$p_{\bag\doc} \in[ 0, 1]$ & the probability of an instance $\mathbf{x}_{\bag\doc}$ in bag $\mathcal{X}_{\bag}$ in super bag to be positive\\
\hline
$C \in \{1,2,...,K\}$ & multi-class label of super bag\\
\bottomrule
    \end{tabular}%
\end{table}%

Given, a collection of streaming 
media sources (e.g., news feeds, blogs and social network streams), the 
objective 
of our study is to develop a machine learning approach 
to forecast the occurrence 
of an event of interest in the near future. Specifically, 
we focus on forecasting ``protests'' or civil unrest movements in Latin America  from a daily collection 
of published news articles.  Besides forecasting the 
protest, we want to identify 
the specific news articles from the streaming news outlets that can be considered as 
supporting evidence for 
further introspection by an intelligence 
analyst. We refer 
to these identified articles as 
\emph{precursors} for a specific protest. 

\begin{figure}[t]
\centering
\includegraphics[width=0.8\textwidth]{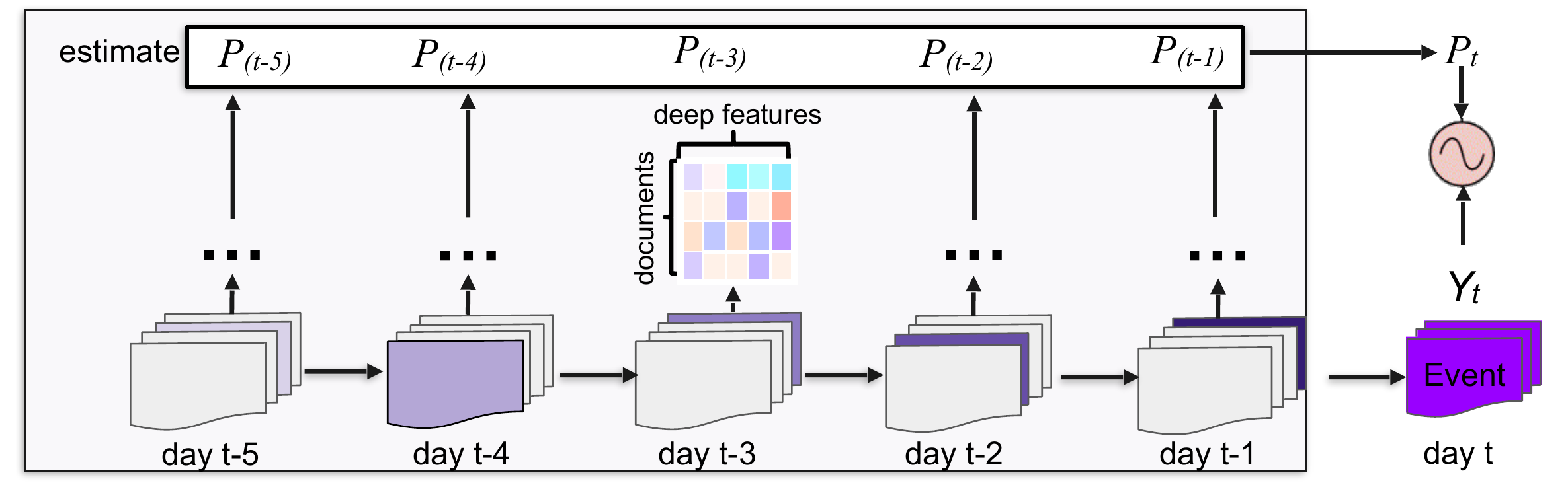}
\caption{Overview of Proposed Approach for Forecasting and Precursor Discovery.}\label{fig:overview}
\end{figure}

Figure \ref{fig:overview} provides an overview of our proposed approach and 
problem formulation. We show groups of news articles collected daily, five days prior 
to the specific protest event (being forecast). Within our proposed MIL-based 
formulation, each news article 
is an individual instance, the collection of news articles published on a given day 
is a bag, and the ordered collection of bags (days) is denoted by super-bag (explained in detail later). 
For this study, each 
individual news article is represented by a distributed representation for text derived using deep learning 
framework and text embeddings \cite{DBLP:LeM14}. 
Figure \ref{fig:overview} shows that for certain days within the collection we attempt to
identify 
news articles (highlighted) that 
are considered as precursors from  the entire collection of input news articles  used for 
forecasting the occurrence of a specific target. 

\subsection{Formal Definition and Notations}

For a given protest event \evt occurring on day $t+\lead$, we 
assume that for each day before the event 
we are tracking a multitude of news 
sources. We 
represent the 
collection of $n_\bag$ news articles published on a 
given day $\bag$ by $\mathcal{X}_\bag = \{\bf{x}_{\bag,1} \ldots \bf{x}_{\bag,n_\bag}\}$, 
where 
the $\doc$-th  news article is represented by $\bf{x}_{\bag\doc}$.
The ordered collection of news articles for the protest event
up to day $t$ can be represented as a super-bag, 
$\mathbb{S}_{1:t} =\{\mathcal{X}_1, \ldots, \mathcal{X}_t\}$.
The occurrence of the protest event at time $t+\lead$ is denoted by $\mathcal{Y}_{t+\lead} \in \{-1,+1\}$ where $1$ denotes a 
protest and $-1$,
otherwise. 

The forecasting problem can be formulated as learning a mathematical 
function $f(\mathbb{S}_{1:t}) \rightarrow \mathcal{Y}_{t+\lead}$
that maps the input, ordered collection of news 
articles extracted per day to a protest 
indicator $\lead$ days in the future from 
the day $t$. 
To identify the news articles considered as 
precursors (evidence), we want to estimate 
a probability  for each news article on any given day that signifies 
the occurrence of a given protest. For a news article 
$\bf{x}_{\bag\doc}$, we denote this estimated probability 
value by $p_{\bag\doc}$. As such, 
given the collection of news articles we identify the precursor set as the ones 
with $p_{\bag\doc}$ greater than a fixed threshold $\tau$. We  represent this 
precursor set of 
documents as a subset of the original super-bag, given by $\{{\bf{x}_{\bag\doc}}  \in \mathbb{S}_{1:t} \mid  p_{ij} > \tau\}$.
As a secondary objective, we want  
to forecast the occurrence of an event with a 
long lead time i.e., large values of $\lead$. 
Table \ref{tab:notations} captures the notations and definitions used in this study.

\section{Methods}\label{sec:method}

We first provide our 
intuition behind formulating the precursor discovery and 
forecasting problem within a novel extension of multiple 
instance learning algorithm. Parallel to the 
standard multiple instance learning algorithms we have a 
group of news articles (bags) with labels available only for the 
entire bag (i.e., leading to a protest); and one of the objectives 
is to train a classifier to predict the bag-level label.
In addition to predicting the group-level labels, we also care about predicting the labels 
for individual news articles (instances) since, they signify the precursor. Various MIL 
  formulations extend this approach to estimate the key instances within a bag or 
  provide instance-level labels. However, our problem setting has a two-level grouping structure with 
  sequential constraints i.e., we capture news articles per day (bags) and group the days to form a super-bag with 
  labels only available at the super-bag level. As such, we propose nested multiple instance learning formulations 
  for predicting the super bag level labels (forecast), and then estimate the 
    bag-level and instance-level probabilities for identifying association of the bag and instance with the event, respectively. 
We developed 
various extensions of our proposed approach, that 
tie the different sequential and group constraints.

\subsection{Nested MIL model~(\nmil)}

We model the instance level probability estimates  $p_{\bag\doc}$ for a news article 
$\doc$ on day $\bag$ to associate with a targeted event \evt  
with a logistic function. This
probability estimates indicate how related the specific instance is to the
target event, \evt. Higher the probability value, the more related the document is to
the target event and most probably represents a precursor  that contains information about causes of 
the 
target event.

\begin{align}
p_{\bag\doc}  = \sigma(\mathbf{w}^T\mathbf{x}_{\bag\doc})= \frac{1}{1 + e^{-\mathbf{w}^T\mathbf{x}_{\bag\doc}}}.
\end{align}

Here, $\mathbf{w}$ denotes the learned weight vector for our model. 
The probability for a day (or bag)  is then modeled as the average of probability estimates
of all instances in a day~\cite{Kotzias:2015:KDD}. Hence, for each bag:
\begin{align}
\mathbb{P}_{\bag}= \mathcal{A}(\mathcal{X}_{\bag}, \mathbf{w}) = \frac{1}{n_{\bag}} \sum_\doc^{n_{\bag}} p_{\bag\doc},
\end{align}
where $\mathcal{A}$ is an aggregation function. 

We then model the probability of 
a super-bag $\mathbb{S}$ (associated with an event \evt) being positive as the
average of the probability of all $t$ bags
within the super bag to be
positive (related to the target event).
Thus:
\begin{align}
P= \mathcal{A}(\mathbb{S}, \mathbf{w}) = \frac{1}{t} \sum_\bag^{t} \mathbb{P}_{\bag}
\end{align}

For a given super bag $\mathbb{S}$, as all the $t$ 
bags within it are
temporally ordered, the probability estimates  for a given bag (day) is assumed to be similar 
to its 
immediate predecessor. This consistency in consecutive bag
probabilities is modeled by minimizing the following cross-bag cost as below:

\begin{align}
	g(\mathcal{X}_{\bag},\mathcal{X}_{\bag-1})=  (P_{\bag}- P_{\bag-1})^2          
	\label{eq:crossbag_cost_nmil}
\end{align}

Finally, given a set of true labels $Y$ for the super bags, we can train our 
model by minimizing the following cost function w.r.t to $\mathbf{w}$: 



\begin{align}  
J(\mathbf{w}) = & \frac{\beta}{n}\sum_{\mathbb{S}\in \mathcal{S}} f(\mathbb{S}, Y, \mathbf{w}) + \frac{1}{n} \sum_{\substack{\mathbb{S}\in \mathcal{S};\\\mathcal{X}_\bag,\mathcal{X}_{\bag-1}\in\mathbb{S}}} \frac{1}{t} \sum_{\bag=1}^{t}g(\mathcal{X}_{\bag}, \mathcal{X}_{\bag-1}, \mathbf{w})\\ \nonumber
& + \,\frac{1}{n} \sum_{\substack{\mathbb{S}\in \mathcal{S};\mathcal{X}_\bag\in\mathbb{S}\\\mathbf{x}_{\bag\doc}\in\mathcal{X}_\bag}} \frac{1}{t} \sum_{\bag=1}^{t} \frac{1}{n_{\bag}}\sum_{\doc=1}^{n_{\bag}}h(\mathbf{x}_{\bag\doc}, \mathbf{w})+\,\lambda R(\mathbf{w})\, \label{eq:cost-model}
\end{align}

Here,

$\bullet$ $f(\mathbb{S}, Y, \mathbf{w}) = - \mathbf{I}(Y=1) logP_i - \mathbf{I}(Y =-1)(log(1-P))$ is the negative log-likelihood function that penalizes the difference between prediction
and the true label for super bag $\mathbb{S}$ where $\mathbf{I}(\cdot)$  is the indicator function.

$\bullet$ $g(\mathcal{X}_{\bag}, \mathcal{X}_{\bag-1}, \mathbf{w})$ is the
cross-bag cost defined in Equation.~\ref{eq:crossbag_cost_nmil}


$\bullet$ $h(\mathbf{x}_{\bag\doc},\mathbf{w})= \textit{max}(0, m_0 - \textit{sgn}(p_{\bag\doc}- p_0)\mathbf{w}^T \mathbf{x}_{\bag\doc})$
  represents the instance level cost. Here, $\textit{sgn}$ is the sign
  function; $m_0$ is a crucial margin parameter used to separate the
  positive and negative instances from the hyper line in the feature
  space; $p_0$ is a threshold parameter to determine positiveness of
  instance.

$\bullet$ $R(\mathbf{w})$ is the regularization function.

$\bullet$ $\beta$, $\lambda$ are constants that control the trade-offs between the loss function and regularization function.



\subsubsection{Cross-bag Similarity (\nmildelta)}

The cross-bag similarity $g(,)$ in the above equation does not allow 
for sudden changes in the day-level probabilities
caused due to newer events happening on the current day.
We update the  cost function across days (bags) (Equation~\ref{eq:crossbag_cost_nmil}) as follows:
\begin{align}
g(\mathcal{X}_{\bag}, \mathcal{X}_{\bag-1}) =  \Delta(\mathcal{X}_{\bag},\mathcal{X}_{\bag-1})(P_{\bag}- P_{\bag-1})^2 
\end{align}
The objective function above allows for label information to spread over the manifold in the feature-space. As such, we 
compute $\Delta (,)$ as the pairwise cosine similarity between the news articles in 
$\mathcal{X}_{\bag}$ and $\mathcal{X}_{\bag-1}$. Since, we do not have ground truth labels for the bag level (day) we 
make this consistency assumption that estimated probabilities for consecutive days should be similar if the 
news articles have similarity in 
the feature space as well. This model is referred by \nmildelta and allows  for sudden changes in how events unfold.

\subsection{Sequential Model (\nmilseq)}

The basic \nmil models assume that there exists a single weight vector across all the days (bags) within a 
super bag. To model the sequential characteristics of the  articles published across consecutive days,  we extend 
this formulation by learning individual weight vectors for each of the historical days. Assuming $t$ days within 
a super bag $\mathbb{S}$ we learn a weight vector for each individual day represented 
as $\Omega = [\mathbf{w}_1,\ldots,\mathbf{w}_t]$; where $\mathbf{w}_j$ is 
the weight vector learned for day $j$. In this setting, the individual weight vectors 
are still learned together in a joint fashion as the Multiple-Task Learning approaches~\cite{Caruana:19972}. 
%
%
However, the probability of a news 
article $\doc$ on day $\bag$ will be given by $p_{\bag\doc} = \sigma(\mathbf{w}_\bag^T\mathbf{x}_{\bag\doc})$. 
The is formulation  is called \nmilseq and given by:

\begin{align}  
J(\mathbf{\Omega}) = & \underbrace{\frac{\beta}{n}\sum_{\mathbb{S}\in \mathcal{S}} f(\mathbb{S}, \mathbf{\Omega}, Y)}_{\text{empirical loss}} + \underbrace{\frac{1}{n} \sum_{\substack{\mathbb{S}\in \mathcal{S};\\\mathcal{X}_\bag,\mathcal{X}_{\bag-1}\in\mathbb{S}}} \frac{1}{t} \sum_{\bag=1}^{t}g(\mathcal{X}_{\bag},\mathcal{X}_{\bag-1},\mathbf{w}_\bag)}_{\text{sequential loss}}\\ \nonumber
& + \underbrace{\frac{1}{n} \sum_{\substack{\mathbb{S}\in \mathcal{S};\mathcal{X}_\bag\in\mathbb{S}\\\mathbf{x}_{\bag\doc}\in\mathcal{X}_\bag}} \frac{1}{t} \sum_{\bag=1}^{t} \frac{1}{n_{\bag}}\sum_{\doc=1}^{n_{\bag}}h(\mathbf{x}_{\bag\doc}, \mathbf{w}_\bag)}_{\text{unsupervised loss}}+\,\lambda R(\mathbf{\Omega})\, \label{eq:cost-seq-model}
\end{align}

Just like the Multi-Task learning algorithms, the regularization term $R(\mathbf(\Omega))$ can be modified
to capture the various relationship-based constraints. However, in this study we ignore these specialized 
approaches focusing only on  the MIL paradigm.


\subsection{Multiclass Classification}

We also 
extend our developed $\nmil$ formulations  to solve general purpose 
multiclass classification problems rather than binary classification problems. Within our domain, each labeled 
event is manually attached with two attributes: event type and event population. Event type 
provides information about the nature of the event.
Event population indicates the size/community of people who participated in the protest event.

For the multiclass classification problems, we train one-versus-rest classifiers for each of the classes learning a separate 
weight vector per class. When classifying a super bag to a specific event type/population we first forecast the binary protest
indicator label for a super bag. Next, we apply the multi-class classification only on the predicted positive examples.

\subsection{Optimization}




%

We perform online stochastic gradient decent optimization to solve our 
cost function and test our model on new data to predict super 
bag label. For every  iteration  in our algorithm, we randomly choose a super-bag $(\mathbb{S}, Y)$ 
from the training dataset $\mathcal{S}$ by picking an index $\super \in\{1,\ldots,n\}$ using a
standard uniform distribution. Then we optimize an approximation based on the sampled super-bag by:

\begin{align}
J(\mathbf{w}; \mathbb{S}) = \beta f + \frac{1}{t}\sum_\bag^{t} g_{\bag}+ \frac{1}{t}\sum_\bag^{t}\frac{1}{n_{\bag}}\sum_j^{n_\bag}h_{\bag\doc} + \lambda R(\mathbf{w})
\end{align}
The gradient of the approximate function is given by:  
\begin{align}\label{eq:dw}
&\nabla_tJ(\mathbf{w}) = \frac{\partial J(\mathbf{w}; \mathbb{S})}{\partial \mathbf{w}} = \lambda \mathbf{w} \\ \nonumber
&-\frac{Y-P}{P(1-P)}\frac{1}{t}\sum_\bag^{t}\frac{1}{n_{\bag}}\sum_k^{n_\bag}p_{\bag\doc}(1-p_{\bag\doc})\mathbf{x}_{\bag\doc} \\ \nonumber
&+\frac{1}{t}\sum_\bag^{t}2(P_{\bag}-P_{\bag-1})\frac{1}{n_{\bag}} \sum_\doc^{n_{\bag}}p_{\bag\doc} (1-p_{\bag\doc})\mathbf{x}_{\bag\doc} \\ \nonumber
&-\frac{1}{t}\sum_\bag^{t}2(P_{\bag}-P_{\bag-1})\frac{1}{n_{\bag-1}} \sum_\doc^{n_{\bag-1}}p_{v\doc} (1-p_{v\doc})\mathbf{x}_{v\doc}\\ \nonumber
&-\frac{1}{t}\sum_\bag^{t}\frac{1}{n_{\bag}}\sum_\doc^{n_{\bag}}sgn(p_{\bag\doc} - p0)\mathbf{x}_{\bag\doc}(o_{\bag\doc}) 
\end{align}
where $v=i-1, o_{\bag\doc}=I(sgn(p_{\bag\doc} - p0)\mathbf{w}\mathbf{x}_{\bag\doc}<m0)$.
We update the weight vector using a varied learning rate and $\mathbf{w}_{t+1} = \mathbf{w}_t -\eta_t\nabla_t(\mathbf{w})$ using mini-batch stochastic gradient descent.

\subsection{Precursor discovery using \nmil}\label{precursor}

In the \nmil model, each super-bag consists of an ordered set of 
bags and each bag represents the documents in one day in the 
city for which we are forecasting a protest event.   We present in Algorithm 1  the steps to identify 
news articles as precursors based on their estimated probability given by $p_{\bag\doc} > \tau$. 

\begin{algorithm}
\caption{Precursor Discovery in \nmil}
\begin{algorithmic}[1]
\Procedure{PD-\nmil}{}
\State {\textbf{Input}: $\mathcal{S}=\{(\mathbb{S}_r, Y_r)\}_{r \in n^+},  \mathcal{M}$}
\State {\textbf{Output}: $\{(ps_r, Y_r)\}_{r\in n^+}$}
\For {super bag $(S_r, Y_r)$}
\State {$ps_r = []$}
\For {t = 1,2,...,d(history days)}
\State {$y_{t} = []$}
\For {$k \in \mathcal{X}_{t}$}
\State {$\hat{y}_{tk} = \sigma(\hat{\mathbf{w}}\mathbf{x}_{tk})$}
\If {$\hat{y}_{tk} >\tau$}
\State {$y_{t} \leftarrow (k, \hat{y}_{tk})$}
\EndIf
\EndFor
\State {sort($y_{t}$) by $\hat{y}_{tk}$ in descending order}
\State {$ps_r \leftarrow$ k where k in top($y_{t}$)}
\EndFor
\EndFor
\Return { $\{(ps_r, Y_r)\}_{r\in n^+}$}
\EndProcedure
\end{algorithmic}
\label{algo:pd}
\end{algorithm}

\section{Experiments}\label{sec:experiment}
\subsection{Datasets}

The experimental evaluation was performed on news documents
collected from around 6000 news agencies between July 2012 to December 2014 across 
three countries in South America including Argentina, Brazil and Mexico. 
For Argentina and Mexico, the input news articles were mainly in Spanish and 
for Brazil, the news articles were in Portuguese. 

The ground truth information about  protest events, called gold standard report (GSR) is exclusively
provided by MITRE \cite{Ramakrishnan:2014}. The GSR is a manually created 
list of civil unrest events that happened during 
the period 2012-2014. 
A labeled GSR event provides information about the geographical  location at the 
city level, date, type and population of a civil unrest news report extracted from 
the most influential newspaper outlets within the country of interest. 
These GSR reports are the target events that are used for validation of our forecasting algorithm, and also 
used for analyzing the identified precursors. We have no ground truth available for verifying the validity 
of the precursors. 

\textbf{Argentina}:
We collected data for Argentina from newspaper outlets 
including {\it Clarin} and {\it Lanacion} from the period of 
July 2010 to December 2014. 
There are multiple protest events that happened in Argentina during that time. 
For instance, people protested against the government and utility/electricity-providing 
companies because of heatwaves in Dec. 2013.

\textbf{Brazil}:
For Brazil, we obtained data from news agencies including the three leading
news agencies in Brazil; {\it O Globo, Estadao,} and {\it Jornal do Brasil} from  November 2012 to September 
2013. During this period Brazil faced several mass public demonstrations occurred across several Brazilian cities 
stemming from  a variety of issues ranging from transportation costs, government corruption and 
police brutality. These mass protests were initiated due 
to a local entity advocating for free public transportation. This period had an unusually high social
media activity and news coverage and is also known as
the ``Brazilian Spring"~\footnote{\url{http://abcnews.go.com/ABC_Univision/brazilian-spring-explainer/story?id=19472387}}.

\textbf{Mexico}:
For Mexico, we tracked news agencies including the top outlets: {\it  Jornada, Reforma, Milenio} from January 2013 to December 2014. 
Over 619 days, we noticed 71 news articles per day on average.
There were more than 2000 protest events in this two-year period with major unrest movements in 2013 was led
by teachers and students demanding education reform by protesting against the government. 




\subsection{Experimental Protocol}

The GSR signifies  the occurrence of a protest 
event on  a given day at a specific location. 
To evaluate the MIL-based forecasting and 
precursor discovery algorithms, for each 
protest event we 
extract all the published news articles  
for up to 10 days before 
the occurrence of the specific event. This ordered collection of per-day news documents up to 
the protest day are considered as positive super bags. 
For negative samples, we  identify  consecutive sets
of five days within our studied  time periods for the different countries
when no protest was reported by the GSR. The ordered collection of per-day news documents 
not leading to a protest are considered as negative super bags for the nMIL approach. 
For any news article (i.e., an individual instance) within a positive/negative super-bag we have no label (or ground truth). As part 
of the precursor discovery algorithm, we estimate a probability for an individual instance to signal a protest (by showing evidence). 
It is important to note that the GSR linked news article for a protest is never used for training purposes. 
Having identified the positive and negative samples, we split our datasets into training and testing partitions and perform 
3-fold cross-validation. 

%


We study the performance of forecasting models with varying lead time days and varying historical days. Lead time ($l$) 
indicates the days in advance the model makes predictions and historical days ($h$) is the number of days over which the news 
articles are extracted as input to the prediction algorithms. As an example, if $l$ is set to 1, then the model forecasts if a protest 
event is planned for the next day. Setting the historical days, $h$ to 5 denotes that we use news from five days before the current day to make 
the forecast. We varied $l$ from 1 to 5 and $h$ from 1 to 10 and trained 50 different models for the different approaches to study 
the characteristics of the developed approaches with varying lead time and historical days.

 
%

For event forecasting, we evaluate the performance by the standard metrics including precision, recall, accuracy and F1-measure. 

\subsection{Comparative Approaches}
We compare the proposed \nmil models to the following approaches:
%
\begin{itemize}
\item \svm: We use the standard support vector machine formulation \cite{Cortes:1995} by collapsing the 
nested grouping structure and  assigning the same label for each news article as it's  super-bag (for training). 
During the prediction phase, the SVM yields the final super-bag prediction (forecast) 
by averaging the predicted label 
obtained for each of the  instances. 
%
%
%

\item \misvm~\cite{andrews2002support}: The MI-SVM model extends the notion of a margin from individual patterns to bags.
  Notice that for a positive bag the margin is defined by the margin of
  the ``most positive'' instance, while the margin of a negative bag is
  defined by the ``least negative'' instance.  In our case, we collapse the news articles 
 from the different historical days into one bag and apply this standard MIL formulation. 
\item
  \textbf{Relaxed-MIL}~(\rmilNOR)~\cite{WangZYB15}:~Similar to the MI-SVM baseline, 
  we collapse the news articles into one bag. However, unlike the MI-SVM formulation the \rmilNOR can provide 
  a probabilistic estimate  for a given document within a bag to be positive or negative. 
\item
  \textbf{Modified Relaxed-MIL}~(\rmilAvg):~ This approach is similar to the \rmilNOR, except we  compute the probability of a
  bag being positive by taking average of estimate of each instance in
  the bag rather than using the Noisy-OR model discussed above.


\item \gicf~\cite{Kotzias:2015:KDD}:~This model optimizes a 
  cost function which parameterized the whole-part relationship between
  groups and instances and  pushes similar items across different 
  groups to have similar labels. 
\end{itemize}

\subsection{Feature Description}

In practice, finding good feature representations to model the news articles 
is not a trivial problem. 
Traditionally the bag of words representation 
allows for easy interpretation but also needs pre-processing and feature selection.
%
%
Several researchers have developed efficient and effective 
neural network representations for language recently~\cite{Bengio:2003:NPL,DBLP:mccd13,DBLP:MikolovSCCD13}.
%
%
Specifically, we learn deep features for documents  by taking advantage of the existing doc2vec model. 
For each document, we generate a $300$ dimension vector for training with contextual window  size of 
10 in an unsupervised version. We compared the performance of deep features with 
traditional TF-IDF features but the results showed little difference. 
Thus, we only report the evaluation of models with deep features.

\section{Results and Discussion}\label{sec:result}
In this section, we evaluate the performance of the proposed models. Firstly, 
we evaluate the effectiveness and efficiency of the methods on real data in 
comparison with baseline methods on multiple configurations of forecasting tasks.
Then, we study and analyze the quality of precursors with respect to
quantitative and qualitative measures. Multi-class forecasting evaluation is also
provided for one of the countries. At last, we perform a sensitivity analysis of
performance regarding parameters in the proposed model.

\begin{table*}[h]
\small
\centering
\caption{Event forecasting performance comparison based Accuracy (Acc)
and F-1 score w.r.t to state-of-the-art methods. The proposed \nmil, \nmildelta, \nmilseq method outperform  state-of-the-art methods across the three
countries.}
\label{perf_result}
\scalebox{0.9}{
\begin{tabular}{lcccccc}
\toprule
\hline
Method       & \multicolumn{2}{|c|}{Argentina}     & \multicolumn{2}{|c|}{Brazil}     & \multicolumn{2}{|c}{Mexico} \\
\hline
             & Acc   & F-1        & Acc   & F-1       & Acc   & F-1  \\
\hline
\svm         & 0.611($\pm$0.034)      &0.406($\pm$0.072)        & 0.693($\pm$0.040)       &0.598($\pm$0.067)       & 0.844($\pm$0.062)       &0.814($\pm$0.091)  \\
\misvm       & 0.676($\pm$0.026)       &0.659($\pm$0.036)        & 0.693($\pm$0.040)       &0.503($\pm$0.087)       & 0.880($\pm$0.025)       &0.853($\pm$0.040)  \\
\rmilNOR     & 0.330($\pm$0.040)       &0.411($\pm$0.092)        & 0.505($\pm$0.012)       &0.661($\pm$0.018)       & 0.499($\pm$0.009)       &0.655($\pm$0.025)  \\
\rmilAvg     & 0.644($\pm$0.032)       &0.584 ($\pm$0.055)       & 0.509($\pm$0.011)       &0.513($\pm$0.064)       & 0.785($\pm$0.038)       &0.768($\pm$0.064)  \\
\gicf        & 0.589($\pm$0.058)       &0.624($\pm$0.048)        & 0.650($\pm$0.055)       &0.649 ($\pm$0.031)      & 0.770($\pm$0.041)       &0.703($\pm$0.056)  \\
\hline
\nmil        &\emp{0.709}($\pm$0.036)  &0.702($\pm$0.047)        &\emp{0.723}($\pm$0.039)  &0.686 ($\pm$0.055)      &\emp{0.898}($\pm$0.031)  &\emp{0.902}($\pm$0.030)  \\
\nmildelta   & 0.708($\pm$0.039)       &\emp{0.714}($\pm$0.034)  & 0.705($\pm$0.048)       &\emp{0.698}($\pm$0.045) & 0.861($\pm$0.014)       & 0.868($\pm$0.014) \\
\nmilseq &  0.687($\pm$0.038)&0.680($\pm$0.045)  &   0.713($\pm$0.028)   &0.687($\pm$0.038) & 0.871($\pm$0.013) & 0.879($\pm$0.014) \\
\hline
\end{tabular}}
\end{table*}

\subsection{How well does the \nmil  forecast protests?}
\subsubsection{Comparative Evaluation.}

Table~\ref{perf_result} reports the prediction performance of the \nmil 
approach in comparison to other baseline approaches for the task of forecasting protests.
Specifically, we use set $\beta=3.0$, $\lambda=0.05$, $m_0=0.5$ and $p_0=0.5$
and  report the average accuracy and F1 score along with standard deviation 
for predicting protests across multiple runs of varying 
historical days with lead time set to 1. 
%
%
We observe 
that the 
\nmil approaches outperform the  baseline approaches across 
all the three countries. 
The
\rmilNOR approach performs poorly because the 
the noisy-or aggregation function associating 
the bag-level labels to instance-level labels 
forces most of the news articles within the positive bags 
to have probability values close to 1.  However, 
given the large collection of news articles available per day only a 
subset of them will indicate signals/evidence for a protest. 
For Argentina, the  \nmil and \nmildelta approaches 
outperformed the best baseline (\misvm), by \textbf{7\%} and \textbf{8\%} with respect 
the average F1 score, respectively. 


%
%



\begin{figure}[t]
\centering
\begin{subfigure}{0.32\textwidth}
\includegraphics[width=1.0\linewidth]{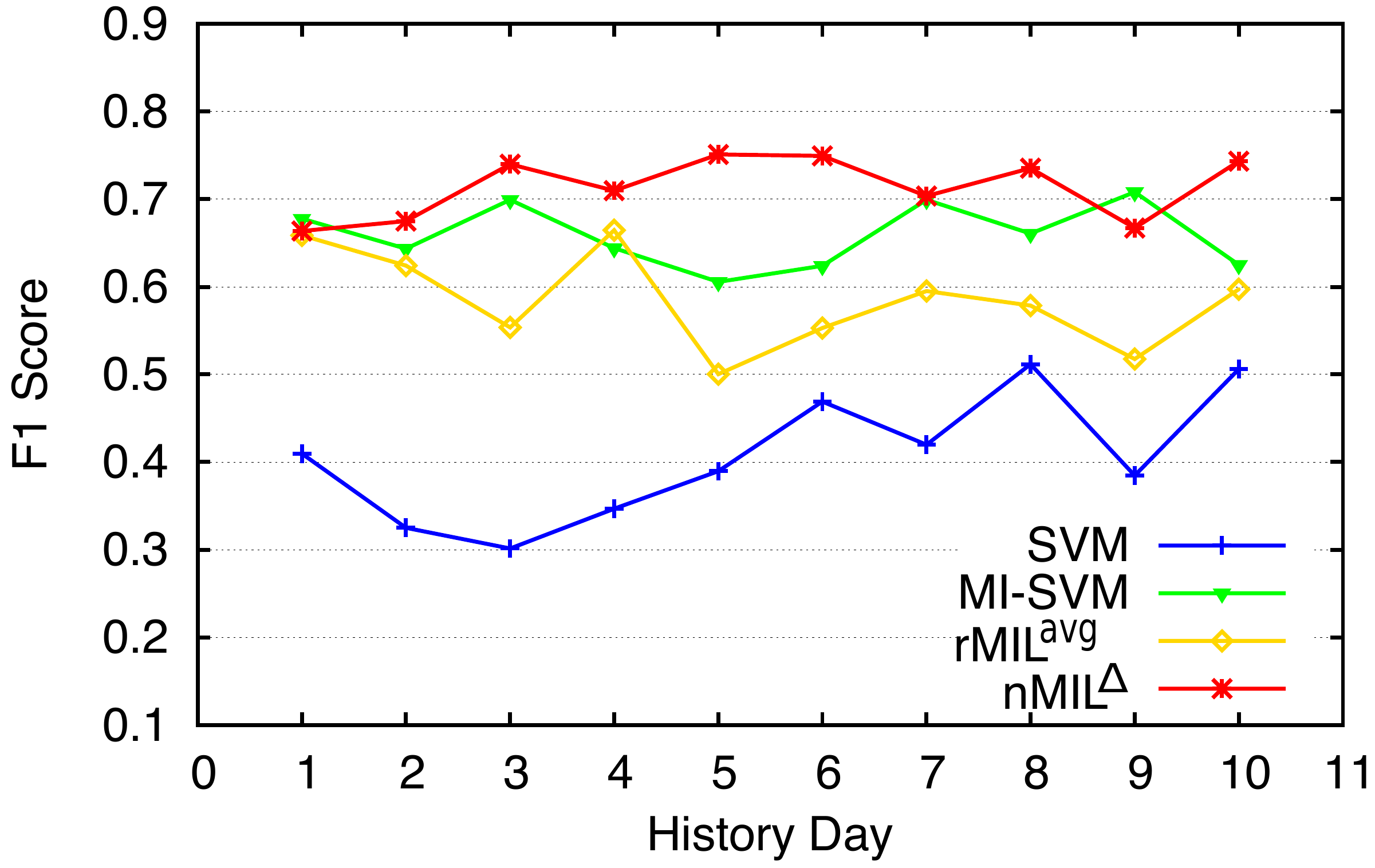} 
\caption{Argentina}
\label{fig:Argentina-F1}
\end{subfigure}
 \begin{subfigure}{0.32\textwidth}
\includegraphics[width=1.0\linewidth]{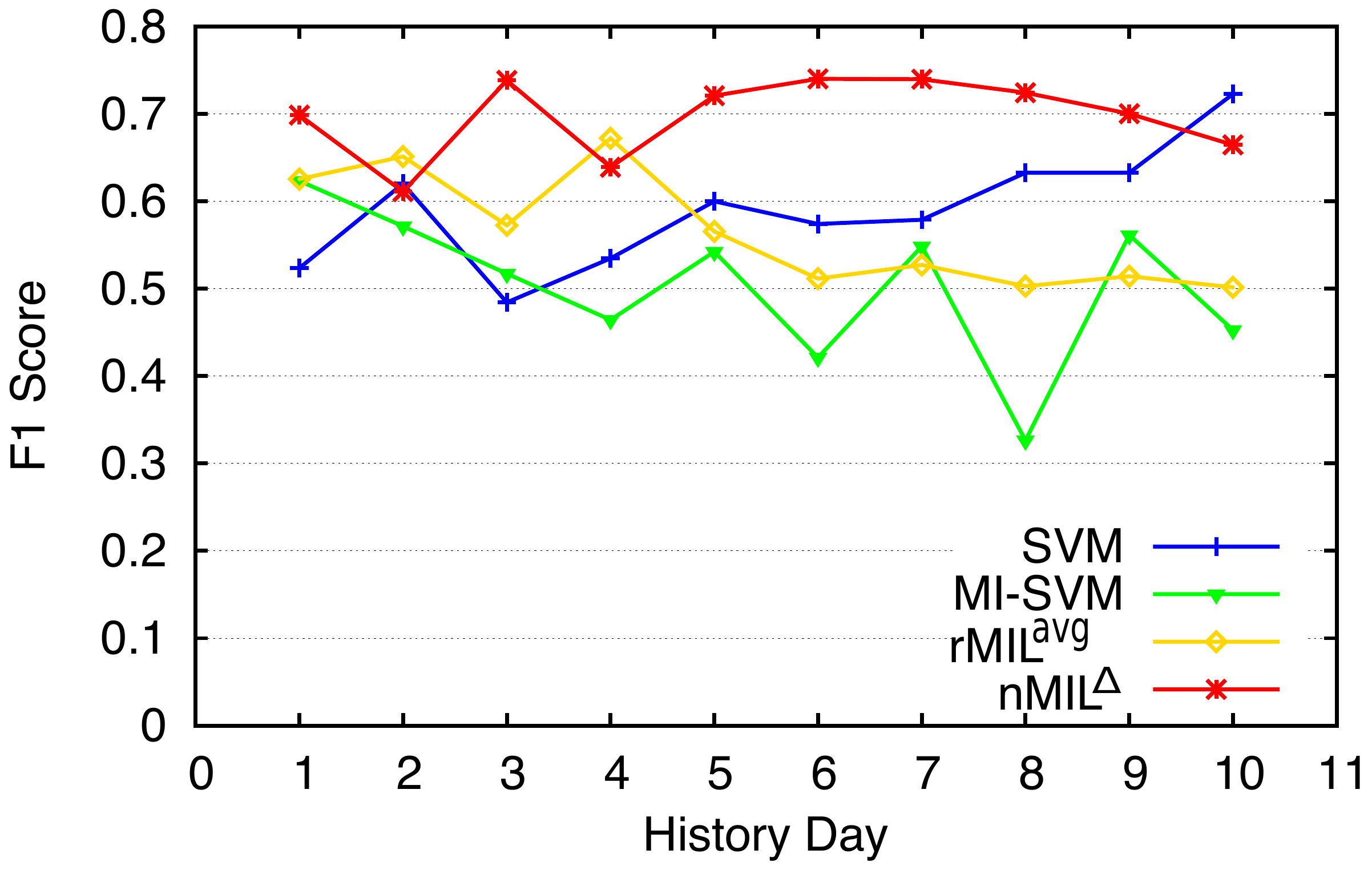}
\caption{Brazil}
\label{fig:Brazil-F1}
\end{subfigure}
\begin{subfigure}{0.32\textwidth}
\includegraphics[width=1.0\linewidth]{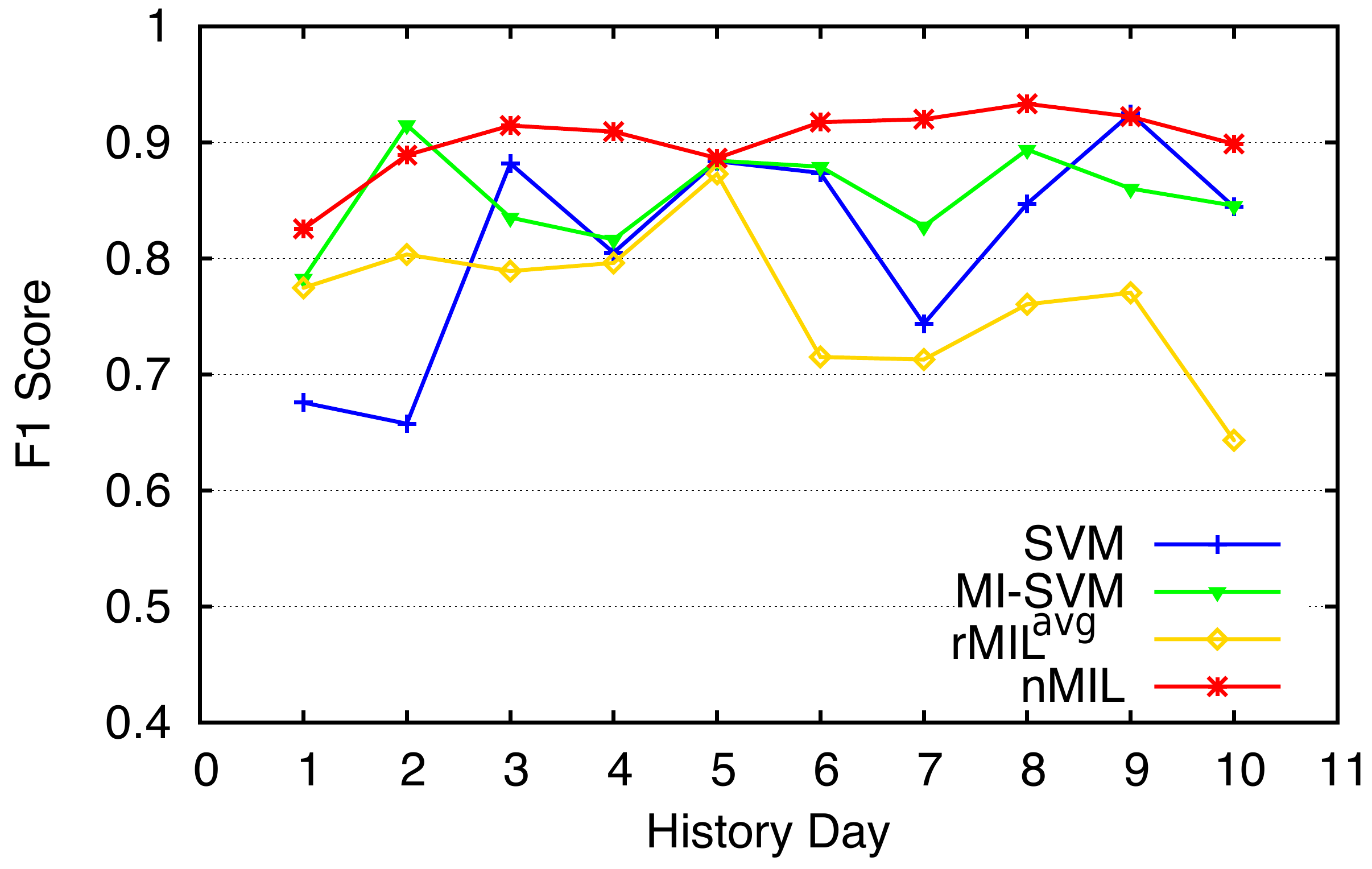}
\caption{Mexico}
\label{fig:Mexico-F1}
\end{subfigure}
\caption{
Forecasting evaluation on 3 countries with respect to F1 score for \svm, \rmilNOR, \rmilAvg and \nmil. X-axis is the number of
historical days used in the training process. Y-axis shows the average F1 score of 10 runs of experiments.}
\label{fig:f1-3c-eval}
\end{figure}

Figure ~\ref{fig:f1-3c-eval} shows the changes to F1 score for the
proposed \nmil approach in comparison to 
\svm, \misvm and 
\rmilAvg for different number of historical days that are used in
training with lead time set to 2. We trained 10 different models that use
different number of historical days respectively varying from 1 to 10. 
These results show the methods that utilize the nested structure (\nmildelta)
within  the multi-instance learning paradigm, generally 
performed better than others. Moreover, the proposed \nmildelta models performed well consistently
across different countries with different number of history days.


%

\subsubsection{How early can the \nmil forecast?}
\begin{table*}[t]
\small
\centering
\caption{F1-measure for \rmilAvg and \nmil models on Argentina, Brazil, and Mexico with history days from 1 to 5.}\label{tab:f1-leadtime-result}
\scalebox{0.6}{
\begin{tabular}{c|c|ccccc|ccccc|ccccc}
\toprule
\hline
                            & Country      & \multicolumn{5}{c}{Argentina}                                                      & \multicolumn{5}{c}{Brazil}                                                         & \multicolumn{5}{c}{Mexico}                                                         \\
\hline                            & History Days & 1              & 2              & 3              & 4              & 5              & 1              & 2              & 3              & 4              & 5              & 1              & 2              & 3              & 4              & 5              \\
\hline\multirow{2}{*}{Leadtime 1} & \rmilAvg      & 0.719          & 0.714          & 0.690          & 0.710          & 0.705          & 0.717          & 0.692          & 0.696          & 0.662          & 0.680          & 0.815          & 0.803          & 0.789          & 0.796          & 0.873          \\
                            & \nmil        & \emp{0.745} & \emp{0.735} & \emp{0.722} & 0.691          & \emp{0.716} & \emp{0.734} & \emp{0.768} & \emp{0.721} & \emp{0.735} & \emp{0.717} & \emp{0.842} & \emp{0.868} & \emp{0.863} & \emp{0.884} & \emp{0.884} \\
\hline\multirow{2}{*}{Leadtime 2} & \rmilAvg     & 0.659          & 0.624          & 0.554          & 0.665          & 0.500          & 0.695          & 0.651          & 0.573          & 0.672          & 0.565          & 0.846          & 0.875          & 0.860          & 0.878          & 0.912          \\
                            & \nmil       & \emp{0.664} & \emp{0.675} & \emp{0.740} & \emp{0.710} & \emp{0.751} & \emp{0.699} & 0.611          & \emp{0.738} & 0.639          & \emp{0.721} & 0.825          & \emp{0.889} & \emp{0.914} & \emp{0.909} & 0.886          \\
\hline\multirow{2}{*}{Leadtime 3} & \rmilAvg      & 0.674          & 0.606          & 0.622          & 0.543          & 0.578          & 0.694          & 0.682          & 0.620          & 0.715          & 0.622          & 0.819          & 0.787          & 0.808          & 0.750          & 0.853          \\
                            & \nmil        & 0.649          & \emp{0.669} & 0.560          & \emp{0.669} & \emp{0.737} & \emp{0.687} & 0.639          & \emp{0.674} & \emp{0.717} & \emp{0.742} & \emp{0.856} & \emp{0.903} & \emp{0.884} & \emp{0.909} & \emp{0.900} \\
\hline\multirow{2}{*}{Leadtime 4} & \rmilAvg      & 0.656          & 0.558          & 0.588          & 0.556          & 0.476          & 0.729          & 0.712          & 0.720          & 0.628          & 0.621          & 0.809          & 0.822          & 0.798          & 0.878          & 0.772          \\
                            & \nmil         & \emp{0.676} & \emp{0.693} & \emp{0.670} & \emp{0.712} & \emp{0.631} & \emp{0.754} & 0.584          & \emp{0.736} & \emp{0.735} & \emp{0.725} & \emp{0.872} & \emp{0.888} & \emp{0.894} & \emp{0.916} & \emp{0.874} \\
\hline\multirow{2}{*}{Leadtime 5} & \rmilAvg      & 0.669          & 0.676          & 0.590          & 0.567          & 0.575          & 0.710          & 0.588          & 0.616          & 0.548          & 0.570          & 0.828          & 0.845          & 0.810          & 0.733          & 0.889          \\
                            & \nmil         & 0.626          & 0.676          & \emp{0.687} & \emp{0.773} & \emp{0.737} & 0.683          & \emp{0.665} & \emp{0.657} & \emp{0.697} & \emp{0.735} & \emp{0.833} & \emp{0.937} & \emp{0.878} & \emp{0.935} & \emp{0.931}
\\
\hline                            
\end{tabular}
}
\end{table*}


In order to study the changes of performance with and without the nested structure, 
we show the F1 score with varying lead times and historical 
days from 1 to 5
for \rmilAvg and \nmil models in Table \ref{tab:f1-leadtime-result}, respectively. 
%
We observe that with larger lead time (i.e., forecasting earlier than later), the \nmil model
does not necessary lose forecasting accuracy; but is sometimes even better. This can be explained by 
the fact that several times protests are planned a few days in advance and 
that civil unrest unfold as a series of 
actions taken by multiple participating entities over a sequence of days. 
As the lead time increases, F1 score for forecasting initially drops and then increases back. This behavior is 
also noted in prior work by  Ramakrishnan et. al ~\cite{Ramakrishnan:2014}, which  includes protest related data from 
these countries.  In comparison to the \nmil model, the \rmilAvg approach, 
which collapses the sequential structure encoded within the history of days
seems to perform inconsistently with increasing lead time.

%

\subsection{Do the precursors tell a story?}


\paragraph*{Quantitative Evaluation}



\begin{figure}[h]
\centering
\begin{subfigure}{0.45\textwidth}
\includegraphics[width=1.0\linewidth]{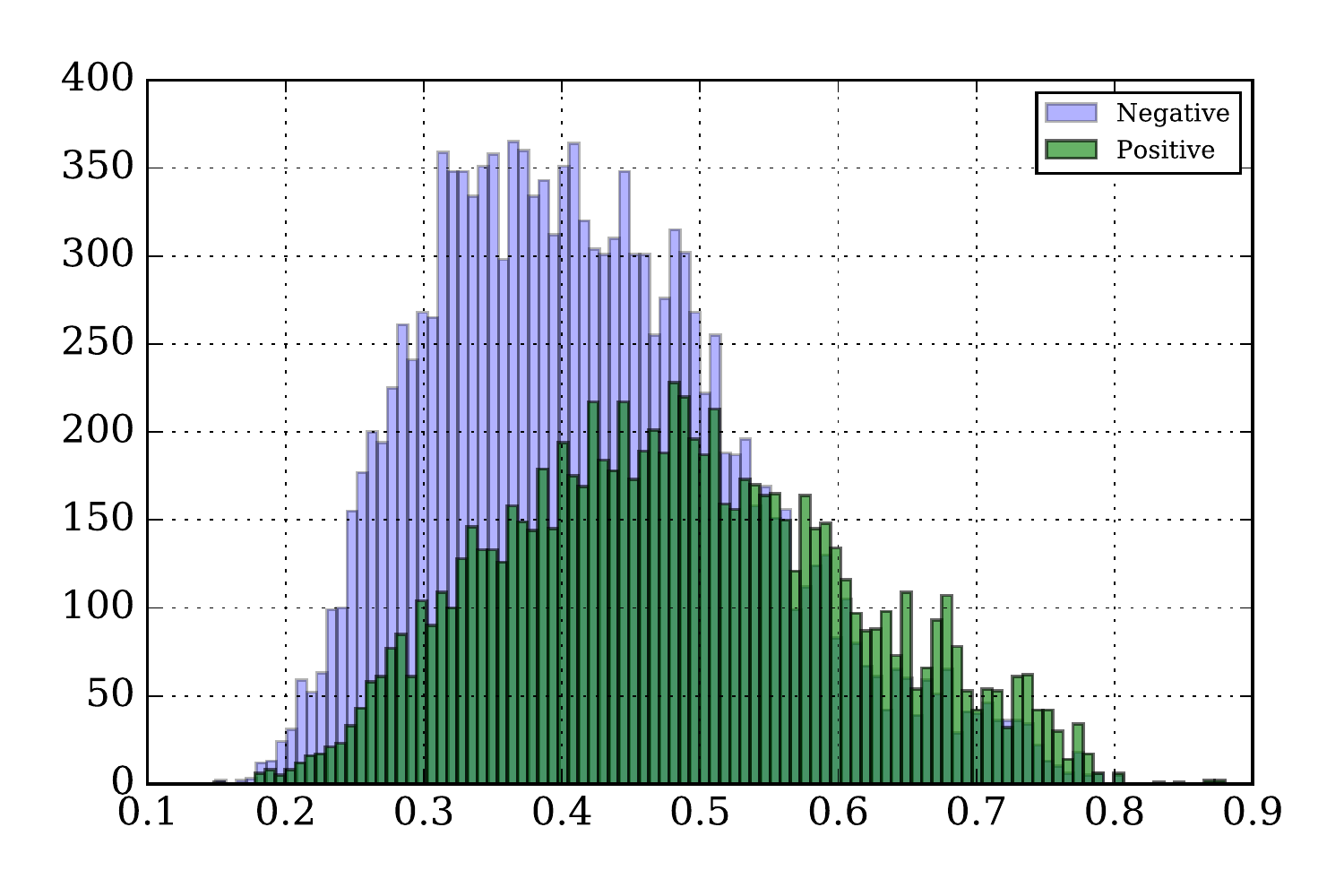}
\caption{Argentina}\label{fig:probs_ar}
\end{subfigure}
\begin{subfigure}{0.45\textwidth}
\includegraphics[width=1.0\linewidth]{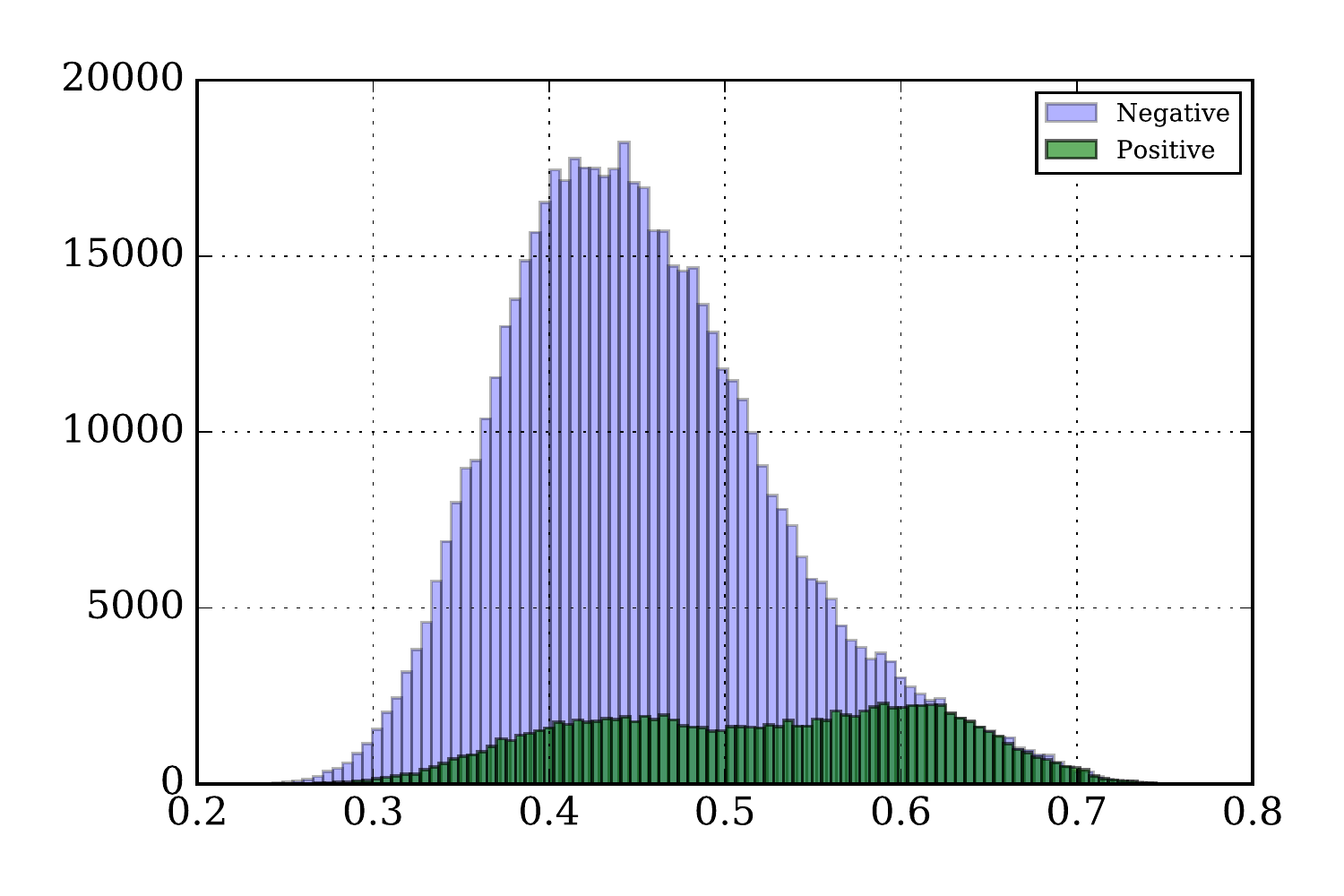}
\caption{Mexico}\label{fig:probs_mx}
\end{subfigure}
\caption{The estimated probabilities for negative examples (purple) and positive examples (green) for Argentina and Mexico}
\end{figure}

Figures \ref{fig:probs_ar} and \ref{fig:probs_mx} show 
the distribution of the estimated probabilities for instances within 
positive and negative super bags for Argentina and Mexico, respectively. 
The  instances within the negative super bags show lower probability estimates  by the proposed model and the 
instances within the positive super bags show higher probability estimates. For Mexico, fewer instances within the positives 
are assigned high probabilities indicating 
strength of the proposed model to identify and rank the precursors. 

\begin{figure}[h]
\centering
\begin{subfigure}{0.45\textwidth}
\includegraphics[width=1.0\linewidth]{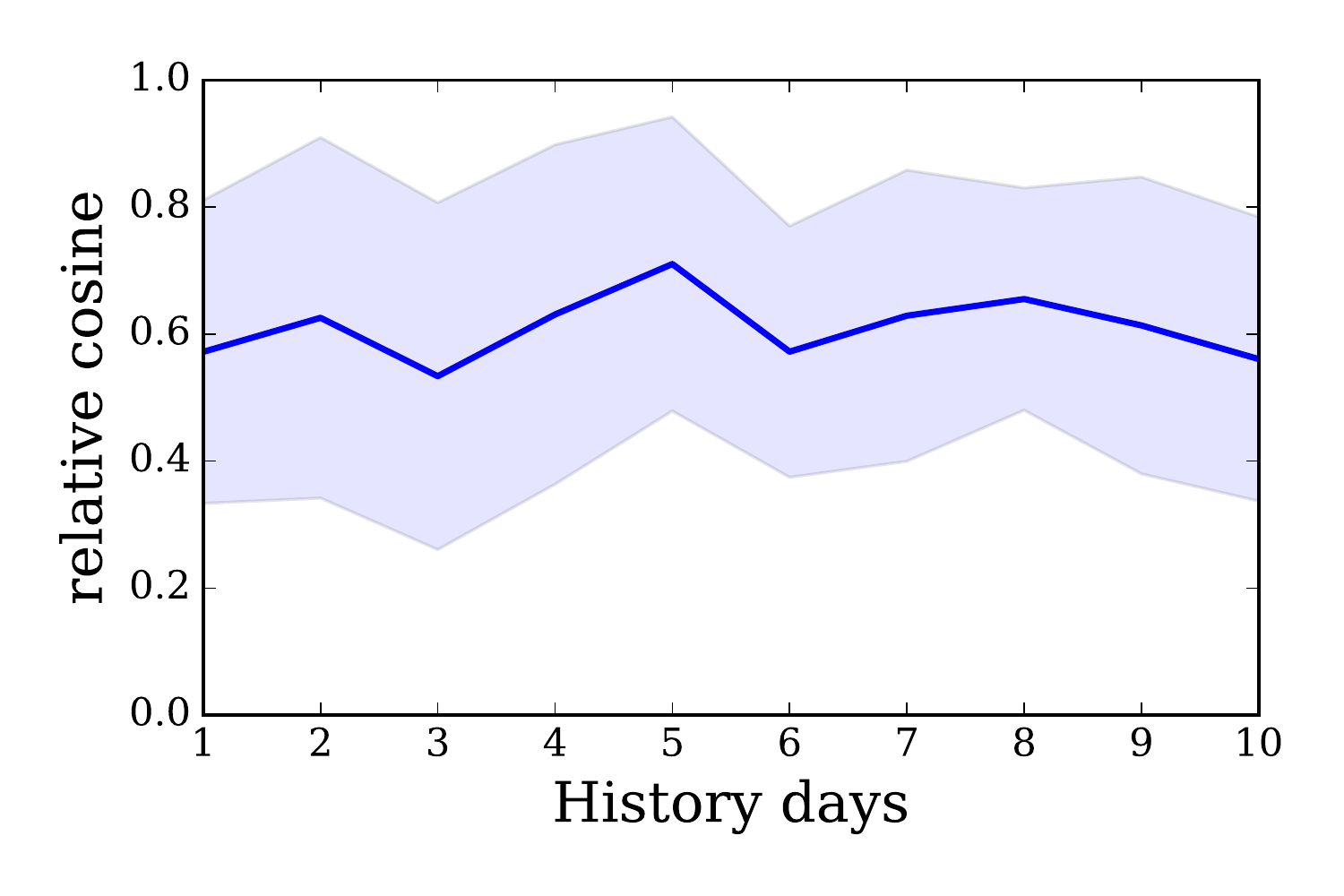}
\caption{Argentina}
\label{fig:cosine-history-ar}
\end{subfigure}
\begin{subfigure}{0.45\textwidth}
\includegraphics[width=1.0\linewidth]{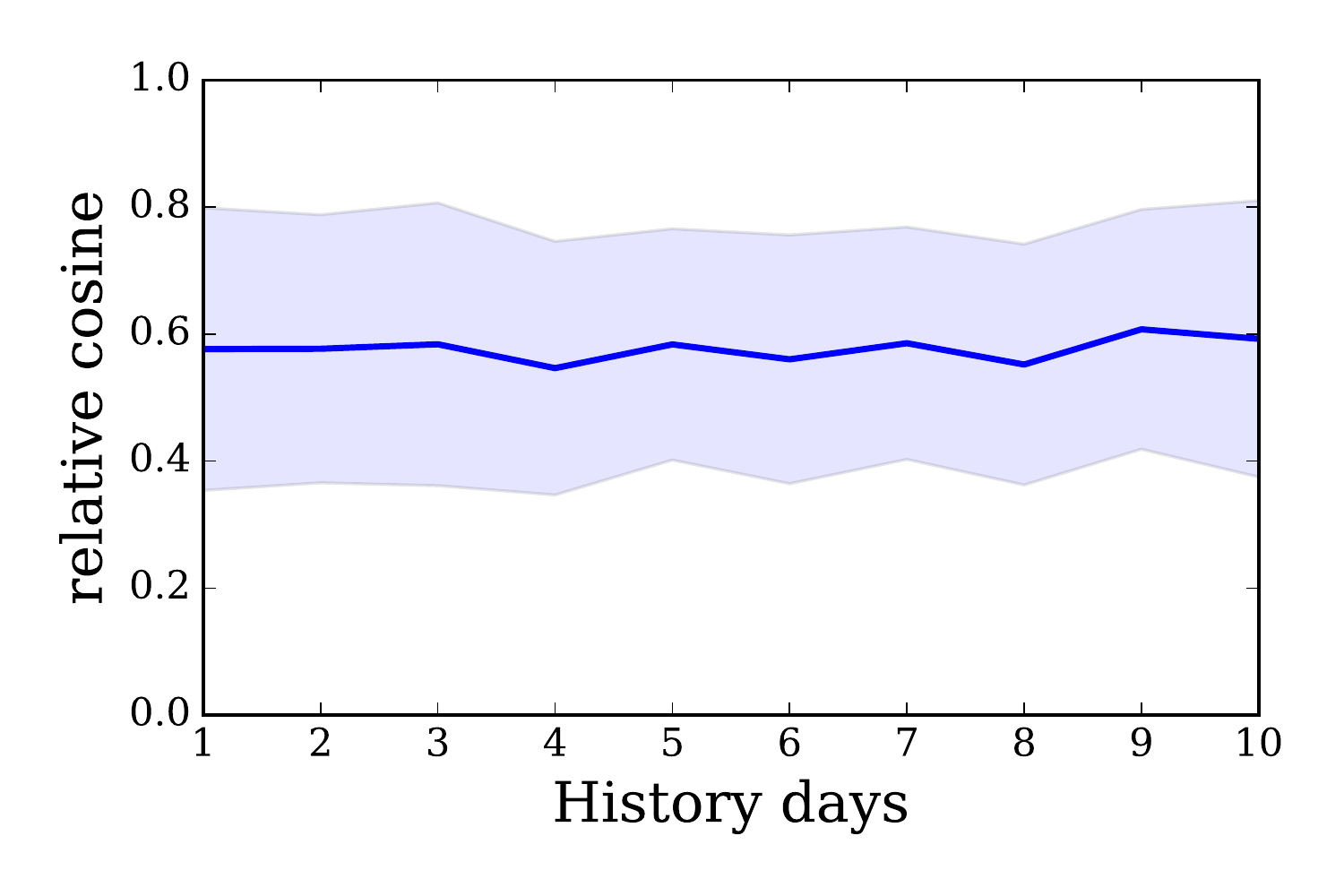}
\caption{Mexico}
\label{fig:cosine-history-mx}
\end{subfigure}
\vspace{-1em}
\caption{Mean of relative cosine values w.r.t target events in history days 
for Argentina and Mexico}
\end{figure}
%
Relative cosine similarity is computed as the pairwise normalized cosine similarity, scaled 
relative to each event.
%
%
Figures~\ref{fig:cosine-history-ar} and~\ref{fig:cosine-history-mx} show 
the average cosine similarity  value for the precursor documents (probability estimate greater than 0.7) with 
the target GSR documents. 
For Argentina, we observe that on average, the documents on 
day 5 have the highest semantic similarity to the target event documents (GSR). The documents 
on day 3 and day 10 have lower similarity compared to the target event.

%

In order to investigate the relationship between the semantic similarity 
and the estimated probability by the proposed models, we 
compare the distribution of relative cosine similarity 
and relative entity hit score of the precursor documents with the target
GSR documents with respect to bag of words 
features. 
Entity words in each news document 
are extracted by an enrichment tool for natural language processing. 
The relative entity hit score is calculated as the the intersection of 
entity set of precursor document and the target event 
divided by the relative minimum length of these two sets.

\begin{figure}[h]
\centering
\begin{subfigure}{0.45\textwidth}
\includegraphics[width=1.0\linewidth]{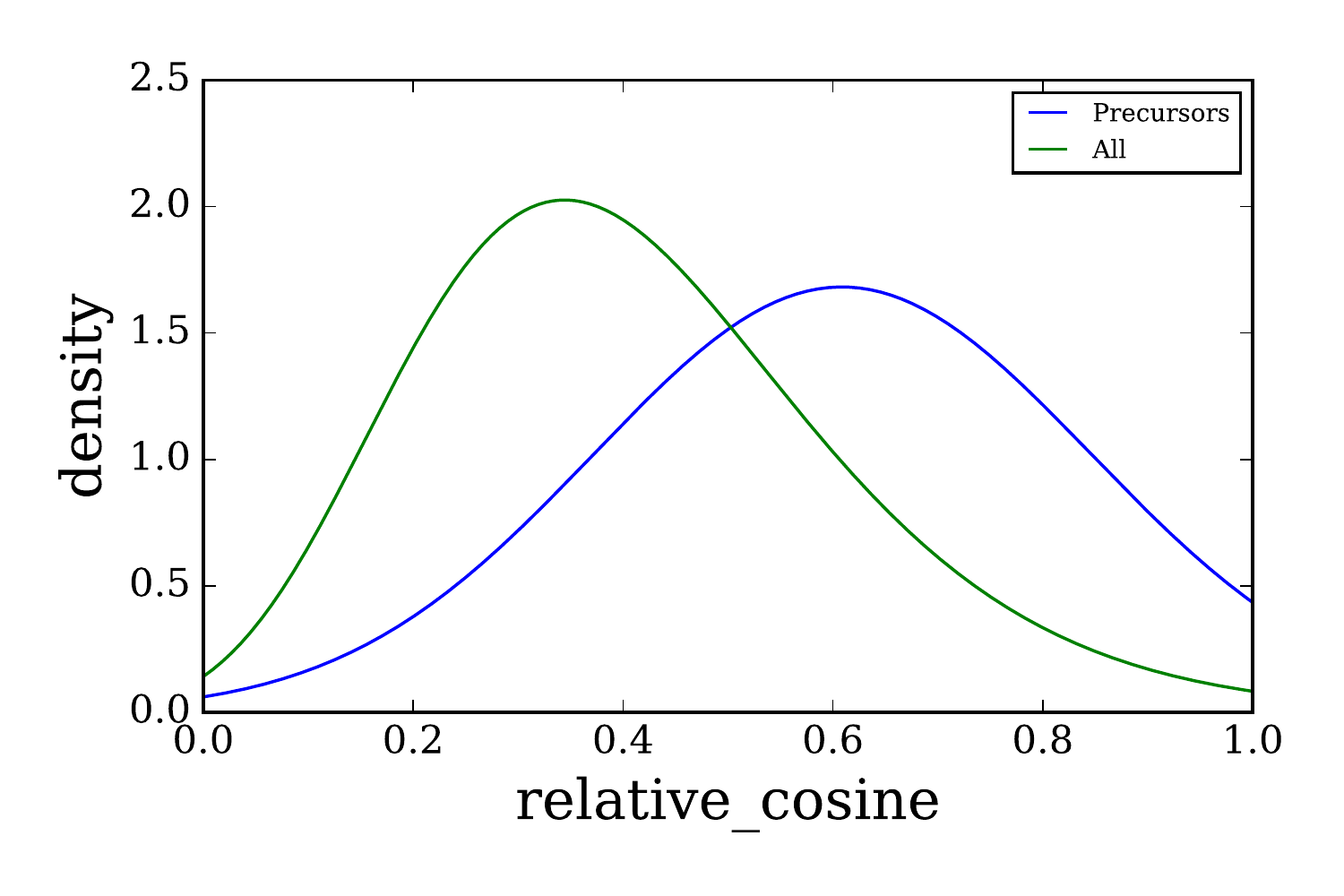} 
\caption{Argentina}\label{fig:cosine_dist_ar}
\end{subfigure}
\begin{subfigure}{0.45\textwidth}
\includegraphics[width=1.0\linewidth]{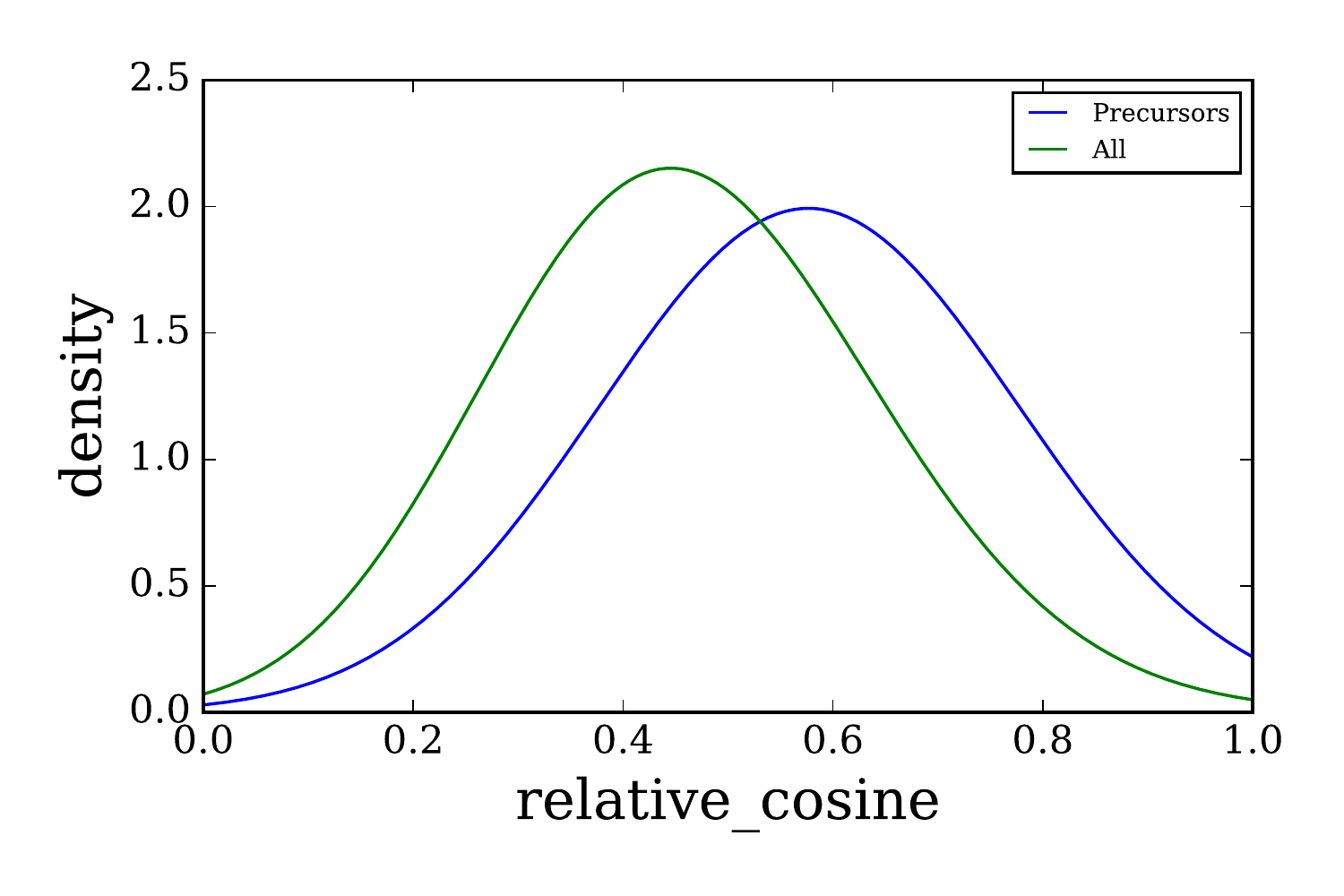} 
\caption{Mexico}
\label{fig:cosine_dist_mx}
\end{subfigure}
\vspace{-1em}
\caption{The figures show the distribution of cosine similarity 
for all documents (green line) and the distribution 
for precursor documents (blue line) with probability greater than $0.7$. 
Left is for Argentina and right is fro Mexico.}
\end{figure}

\begin{figure}[h]
\centering
\begin{subfigure}{0.45\textwidth}
\includegraphics[width=1.0\linewidth]{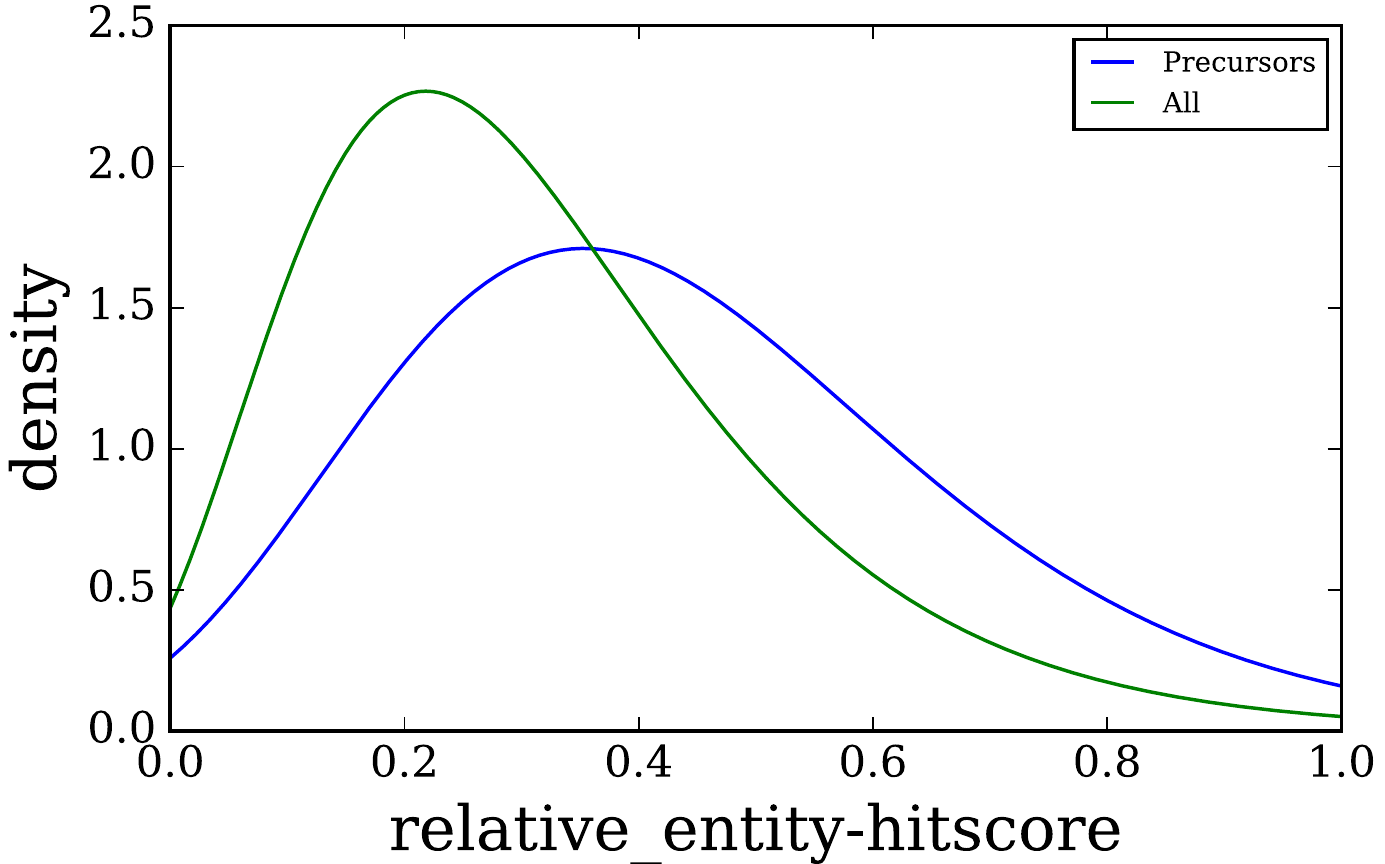} 
\caption{Argentina}
\label{fig:entity_dist_ar}
\end{subfigure}
\begin{subfigure}{0.45\textwidth}
\includegraphics[width=1.0\linewidth]{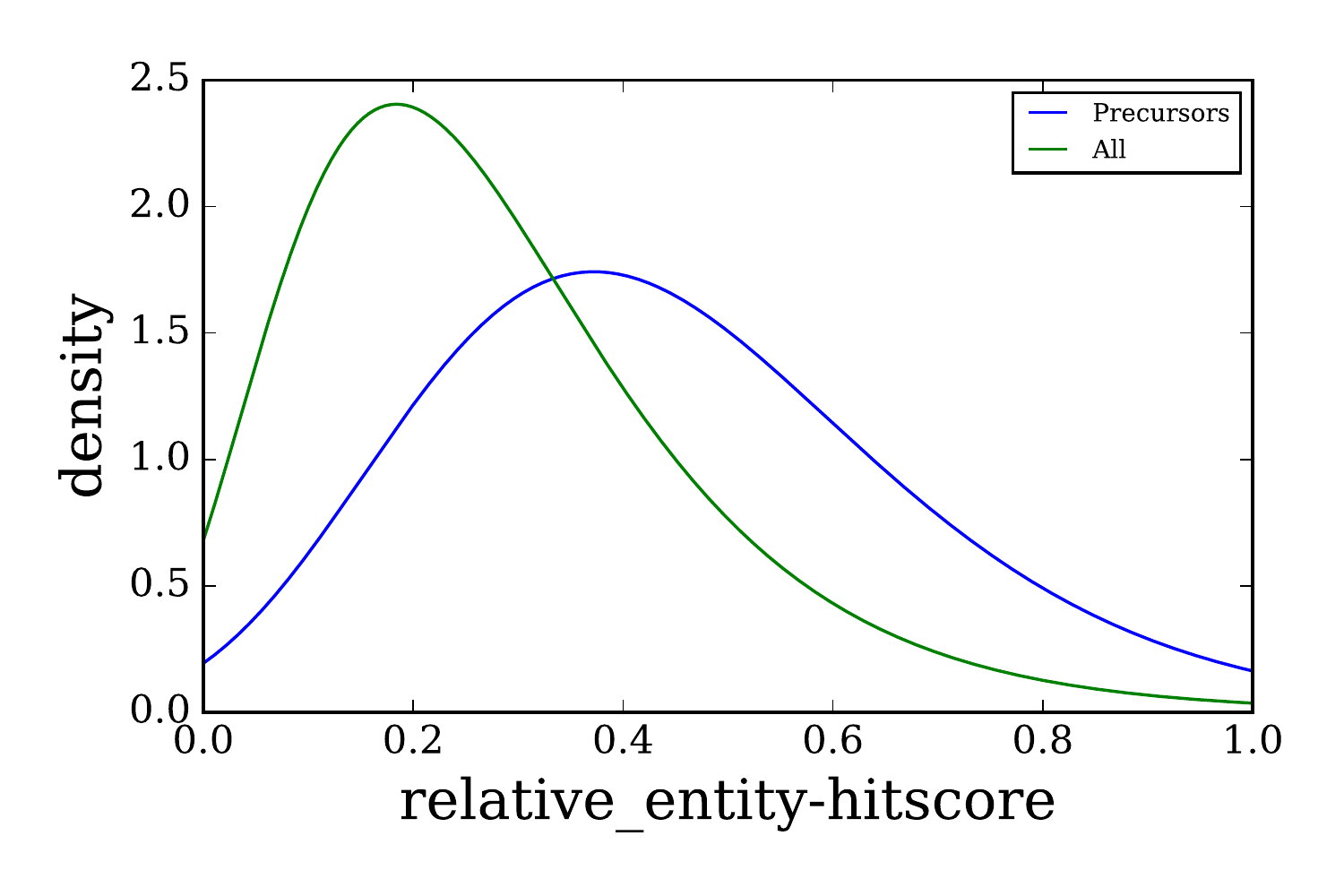} 
\caption{Mexico}
\label{fig:entity_dist_mx}
\end{subfigure}
\vspace{-1em}
\caption{The figures show the distribution of relative entity hit score for all documents (green line) 
and the distribution for precursor documents (blue line) with probability greater than $0.7$. 
Left is for Argentina and right is for Mexico.}
\end{figure}

Figures~\ref{fig:cosine_dist_ar} and~\ref{fig:cosine_dist_mx} show 
the fitted Gaussian distribution of relative cosine similarities 
for all documents (green lines) and precursor documents (blue lines) for Argentina and Mexico, respectively.
Figures~\ref{fig:entity_dist_ar} and~\ref{fig:entity_dist_mx} show the distribution of relative entity hit score for Argentina 
and Mexico, respectively. 
These distribution figures demonstrate that the proposed model 
assigns higher probability to news articles 
with higher semantic similarity 
to the GSR articles representing the protests events. 
These results show the strength of our proposed models in 
identifying the precursor articles. 

%


\paragraph*{Case Studies}

We present  findings about the identified precursors based on the probability estimate by \nmil
across three observed protests. In Figure~\ref{fig:precursor_demo}, we present a protest event against government in Argentina, 
and the selected precursors before its occurrence with their estimated probabilities. 
The titles of news reports as precursors are shown in 
the timeline.

\begin{figure*}[t]
\centering
\begin{subfigure}{0.9\textwidth}
\includegraphics[width=1.0\linewidth]{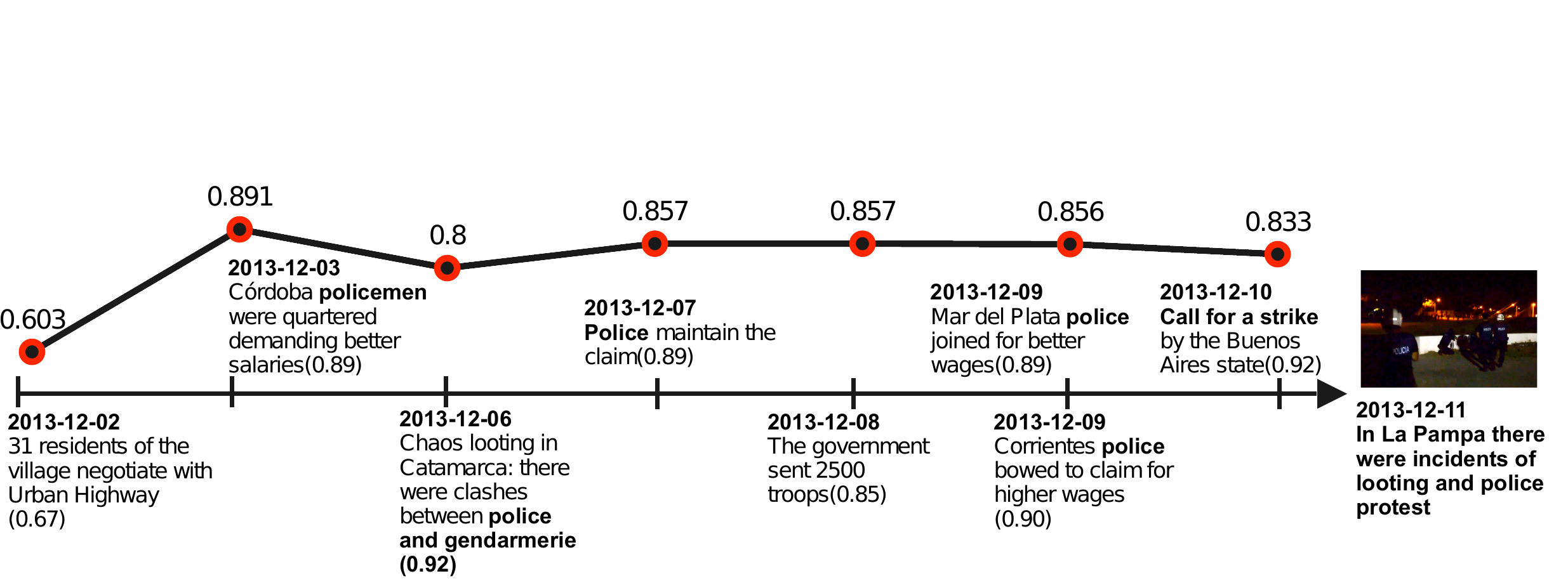} 
\caption{A continuous police protest in Argentina against government for better salary. In the beginning, policemen at Cordoba were requesting for better salaries. Later on, police in Catamarca were  involved in clashes with gendarmerie. Three days before the target event, the government sent out troops and more and more police joined for the same purpose. One day before the event, Buenos Aires state called for a strike.}
\label{fig:ar-precursor-storyline}
\end{subfigure}
\hspace{0.2em}
\begin{subfigure}{0.9\textwidth}
\includegraphics[width=1.0\linewidth]{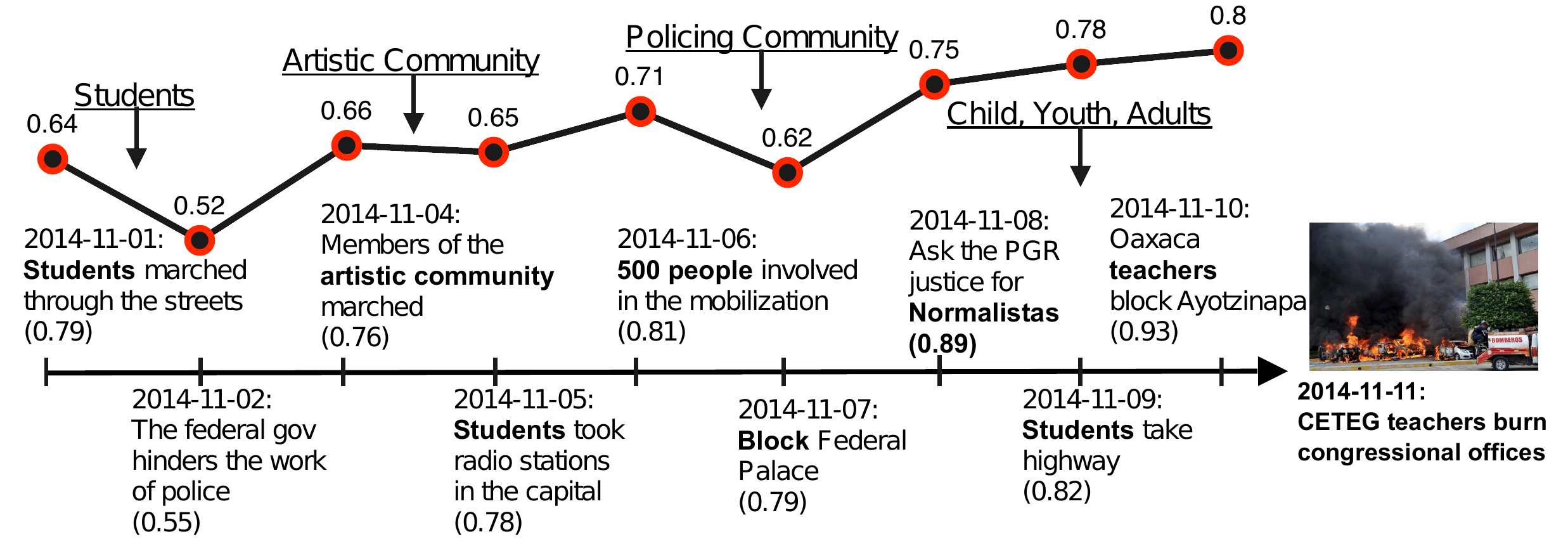} 
\caption{Protester in Mexico burned the congressional offices for justice for the missing teachers. In the beginning, students were marching for justice. Gradually, more communities such as artistic and policing community joined the event. Later on, children, youth, adults, students and teachers blocked traffic for protest.}
\label{fig:mx-precursor-storyline}
\end{subfigure}
\end{figure*}

In Figures~\ref{fig:ar-precursor-storyline} and ~\ref{fig:mx-precursor-storyline}, we present story lines by precursors
that were
discovered for two different protest events in Argentina and Mexico, respectively. Figure~\ref{fig:ar-precursor-storyline} showcases the story line 
about a protest event in Argentina in December 2014. 
In this case, the police were protesting against government for better salaries.
Before this event, clashes between police and gendarmerie (military policy) had occurred leading to involvement of 
several policemen from different parts of the country. 
The text from news articles  demonstrate the tense situation between the police and government in La Pampa, Argentina identified as precursors.


Figure~\ref{fig:mx-precursor-storyline} shows another story line of a continuous protest event 
in Mexico regarding the infamous case of 43 missing 
students~\footnote{\url{https://en.wikipedia.org/wiki/2014_Iguala_mass_kidnapping}}. 
The resulting outrage triggered 
constant protests which were identified by our 
proposed model.  
The figure shows a timeline of how the events 
turned violent leading up to the burning of congressional offices and 
depicts how different  communities joined the movement. 

%
\subsection{Can \nmil forecast event populations?}

We also evaluated the performance of our \nmil approaches for 
predicting the event populations  by solving a multi-class classification problem. 
In Table \ref{tab:f1-mc} we depict the
 weighted-average F1 score for event populations (here,
with categories such as 
\textit{Government, Wages, Energy, Others} drawn from the GSR).
Due to space limitations, we only depict the performance of weighted average F1 score on event population across 1 to 5 historical days with lead time of 1.
\begin{table}[t]
\small
\centering
\caption{Multi-Class F1-Measure for \rmilAvg and \nmil models on Argentina and Mexico with historical days from 1 to 5.}\label{tab:f1-mc}
\scalebox{0.7}{
\begin{tabular}{c|c|ccccc|c}
\toprule
\hline
                                       & \textbf{History Days} & \textbf{1} & \textbf{2} & \textbf{3} & \textbf{4} & \textbf{5} & Average(Variance)\\
\hline\multirow{2}{*}{Argentina}             & \rmilAvg              & 0.512     & 0.512      & 0.473      & 0.417   & 0.457&0.474(1e-3)      \\
                                       & \nmil                  & 0.523      & 0.552      & 0.515      & 0.485      & 0.537 & \textbf{0.524(7e-4)}      \\
\hline\multirow{2}{*}{Mexico} & \rmilAvg              & 0.576      & 0.526      & 0.447      & 0.547      & 0.493&0.518(3e-3)      \\
                                       & \nmil                  & 0.570      & 0.583      & 0.560      & 0.615      & 0.545&\textbf{0.575(7e-4)}   \\
\bottomrule
\end{tabular}
}
\end{table}

The proposed multi-class \nmil model outperforms the multi-class 
\rmilAvg model. On average,  for event population, \nmil outperformed \rmilAvg by $10.5\%$ and $10.6\%$ for Argentina and Mexico, respectively.

\subsection{How sensitive is \nmil to parameters?}
\begin{figure}[h]
\centering
\begin{subfigure}{0.45\textwidth}
\includegraphics[width=1.0\linewidth]{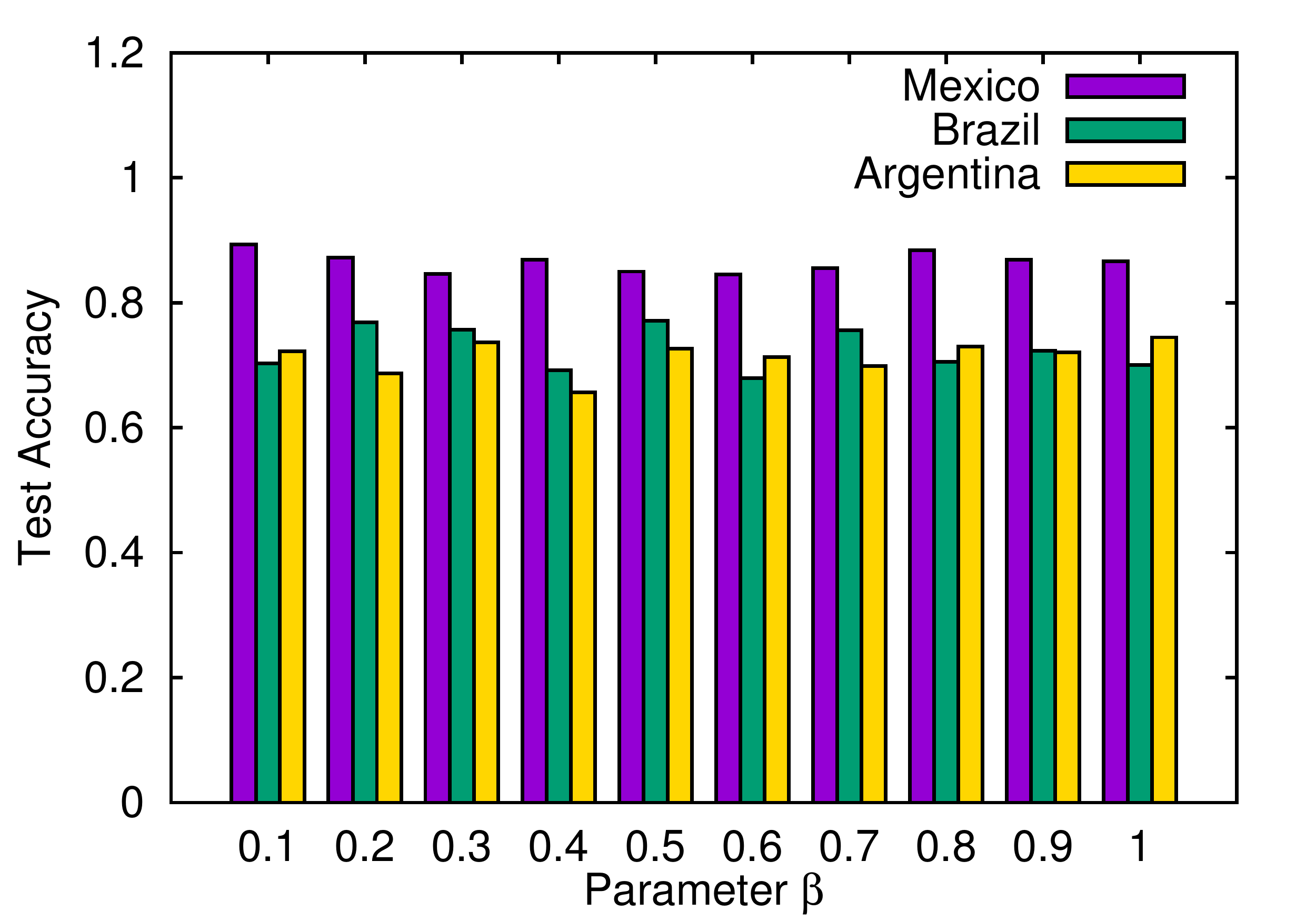} 
\caption{$\beta$ in \nmil}
\label{fig:hrmil-beta}
\end{subfigure}
\hspace{-1em}
\begin{subfigure}{0.45\textwidth}
\includegraphics[width=1.0\linewidth]{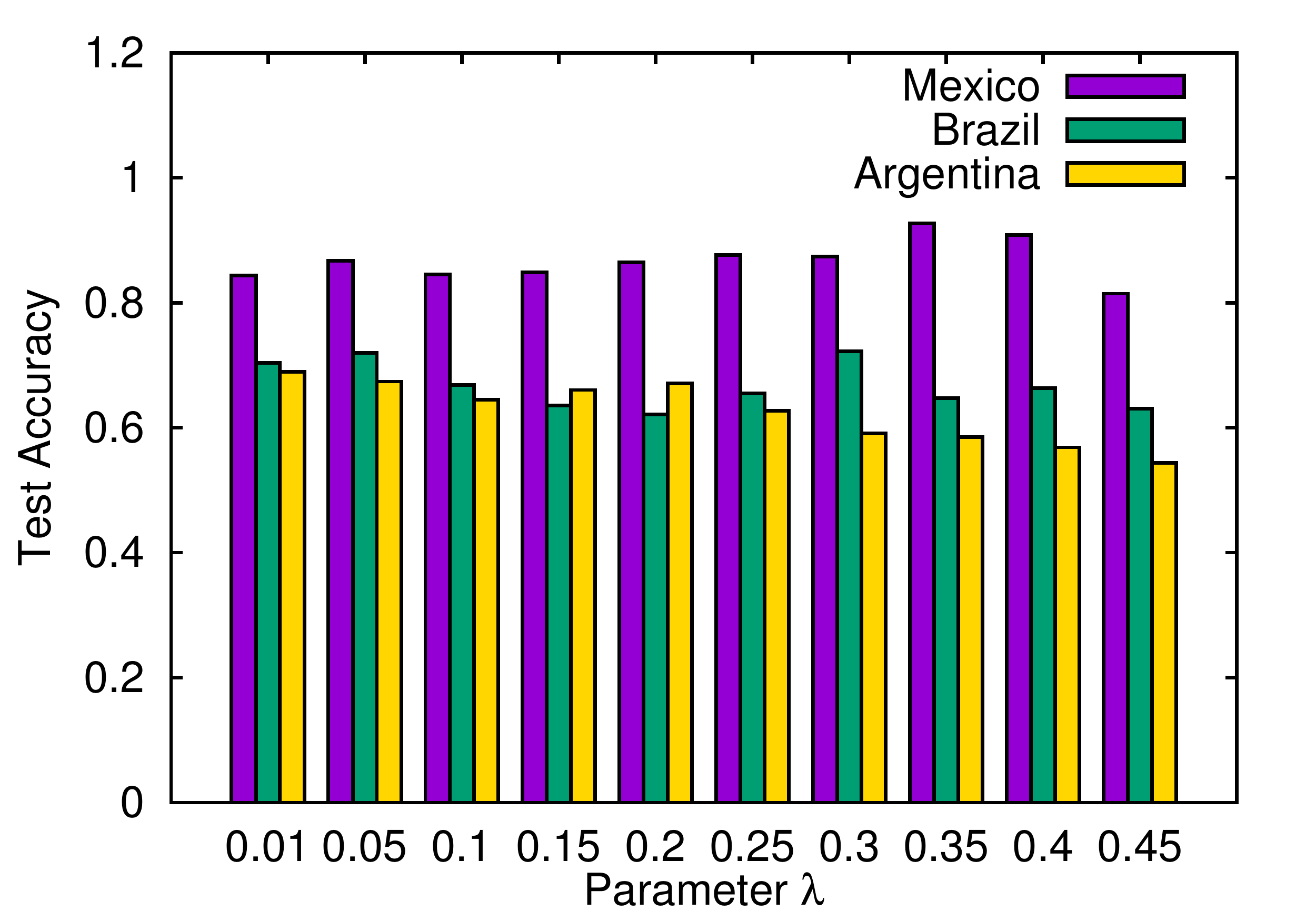} 
\caption{ $\lambda$ in \nmil }
\label{fig:hrmil-lambda}
\end{subfigure}
\vspace{-1em}
\caption{Sensitivity analysis on $\beta$ and $\lambda$. X-axis represents the varying values for the parameter and Y-axis is the test accuracy.}
\end{figure}


There are three main parameters in the proposed \nmil model, 
which are the regularization parameter $\lambda$, weight for super bag loss $\beta$ and
threshold for instance level hinge loss $m_0$.
Figures~\ref{fig:hrmil-beta} and \ref{fig:hrmil-lambda} 
illustrate the performance of the proposed
\nmil by varying $\beta$ and $\lambda$, respectively. 
The test accuracy for different values of $\lambda$ and $\beta$ is 
relatively stable.


\section{Conclusion and Future Directions}\label{sec:conclusion}
This paper presents a novel extension of 
the multi-instance learning framework for event forecasting and 
identifying precursors for protest events. 
Most existing multi-instance approaches 
solve problems in object detection in images, drug activity prediction
or identify sentimental  sentences in text reviews. In contrast, 
we provide a novel application of MIL algorithms that require a two-level 
nested structure for event forecasting and precursor modeling. 

Specifically, we study the strengths of our developed methods on open source news datasets from three 
Latin American countries. Through extensive evaluation and analysis we show the strong forecasting performance 
of the proposed methods with varying lead time and historical data. We also show qualitatively via several case 
studies, the richness of the identified precursors for different protests across different cities.  In the future, we plan to
incorporate heterogeneous data sources like social media streams
for event forecasting within the developed framework.  
We will also extend our nested multi-instance learning framework by exploring regularized multi-task learning approaches 
for enforcing similarity of learned parameters, while enforcing spatial and temporal constraints. 

\section{Acknowledgments}\label{sec:ack}
Supported by the Intelligence Advanced Research Projects Activity (IARPA) via DoI/NBC contract number D12PC000337,
the US Government is authorized to reproduce and distribute reprints of this work for Governmental purposes notwithstanding
any copyright annotation thereon. Disclaimer: The views and conclusions contained herein are those of the
authors and should not be interpreted as necessarily representing the official policies or endorsements, either expressed
or implied, of IARPA, DoI/NBC, or the US Government.


\small
\bibliographystyle{abbrv}
\bibliography{mil_event,mgenomics}

\begin{thebibliography}{10}

\bibitem{Arhrekar:2011}
H.~Achrekar, A.~Gandhe, R.~Lazarus, S.-H. Yu, and B.~Liu.
\newblock Predicting flu trends using twitter data.
\newblock In {\em IEEE Conference on Computer Communications Workshops (INFOCOM
  WKSHPS)}, pages 702--707, April 2011.

\bibitem{amores2013multiple}
J.~Amores.
\newblock Multiple instance classification: Review, taxonomy and comparative
  study.
\newblock {\em Artificial Intelligence}, 201:81--105, 2013.

\bibitem{andrews2002support}
S.~Andrews, I.~Tsochantaridis, and T.~Hofmann.
\newblock Support vector machines for multiple-instance learning.
\newblock In {\em Advances in neural information processing systems}, pages
  561--568, 2002.

\bibitem{Arias:2014}
M.~Arias, A.~Arratia, and R.~Xuriguera.
\newblock Forecasting with twitter data.
\newblock {\em ACM Transactions on Intelligent Systems and Technology (TIST)},
  5(1):8:1--8:24, Jan. 2014.

\bibitem{Bengio:2003:NPL}
Y.~Bengio, R.~Ducharme, P.~Vincent, and C.~Janvin.
\newblock A neural probabilistic language model.
\newblock {\em J. Mach. Learn. Res.}, 3:1137--1155, Mar. 2003.

\bibitem{Bollen2011}
J.~Bollen, H.~Mao, and X.~Zeng.
\newblock Twitter mood predicts the stock market.
\newblock {\em Journal of Computational Science}, 2(1):1 -- 8, 2011.

\bibitem{Caruana:19972}
R.~Caruana.
\newblock Multitask learning.
\newblock {\em Machine Learning}, 28(1):41--75, July 1997.

\bibitem{Cortes:1995}
C.~Cortes and V.~Vapnik.
\newblock Support-vector networks.
\newblock {\em Machine Learning}, 20(3):273--297, Sept. 1995.

\bibitem{Gartner2002}
T.~Gartner, P.~A. Flach, A.~Kowalczyk, and A.~J. Smola.
\newblock Multi-instance kernels.
\newblock In {\em ICML '02: Proceedings of the Nineteenth International
  Conference on Machine Learning}, pages 179--186, San Francisco, CA, USA,
  2002. Morgan Kaufmann Publishers Inc.

\bibitem{He:2013}
J.~He, W.~Shen, P.~Divakaruni, L.~Wynter, and R.~Lawrence.
\newblock Improving traffic prediction with tweet semantics.
\newblock In {\em Proceedings of the Twenty-Third International Joint
  Conference on Artificial Intelligence}, IJCAI, pages 1387--1393, 2013.

\bibitem{Hossain:2012}
M.~S. Hossain, P.~Butler, A.~P. Boedihardjo, and N.~Ramakrishnan.
\newblock Storytelling in entity networks to support intelligence analysts.
\newblock In {\em Proceedings of the 18th ACM SIGKDD International Conference
  on Knowledge Discovery and Data Mining}, KDD, pages 1375--1383, New York, NY,
  USA, 2012.

\bibitem{Kotzias:2015:KDD}
D.~Kotzias, M.~Denil, N.~de~Freitas, and P.~Smyth.
\newblock From group to individual labels using deep features.
\newblock In {\em Proceedings of the 21th ACM SIGKDD International Conference
  on Knowledge Discovery and Data Mining}, KDD, pages 597--606, New York, NY,
  USA, 2015.

\bibitem{Laxman3:2008}
S.~Laxman, V.~Tankasali, and R.~W. White.
\newblock Stream prediction using a generative model based on frequent episodes
  in event sequences.
\newblock In {\em Proceedings of the 14th ACM SIGKDD International Conference
  on Knowledge Discovery and Data Mining}, KDD, pages 453--461, New York, NY,
  USA, 2008.

\bibitem{DBLP:LeM14}
Q.~V. Le and T.~Mikolov.
\newblock Distributed representations of sentences and documents.
\newblock {\em CoRR}, abs/1405.4053, 2014.

\bibitem{jmlr-LiuWZ12}
G.~Liu, J.~Wu, and Z.-H. Zhou.
\newblock Key instance detection in multi-instance learning.
\newblock In {\em ACML}, volume~25 of {\em JMLR Proceedings}, pages 253--268.
  JMLR.org, 2012.

\bibitem{DBLP:mccd13}
T.~Mikolov, K.~Chen, et~al.
\newblock Efficient estimation of word representations in vector space.
\newblock {\em CoRR}, abs/1301.3781, 2013.

\bibitem{DBLP:MikolovSCCD13}
T.~Mikolov, I.~Sutskever, K.~Chen, G.~Corrado, and J.~Dean.
\newblock Distributed representations of words and phrases and their
  compositionality.
\newblock {\em CoRR}, abs/1310.4546, 2013.

\bibitem{OConnorBRS:10}
B.~O'Connor, R.~Balasubramanyan, B.~R. Routledge, and N.~A. Smith.
\newblock From tweets to polls: Linking text sentiment to public opinion time
  series.
\newblock In {\em Proceedings of the Fourth International Conference on Weblogs
  and Social Media (ICWSM)}. The AAAI Press, 2010.

\bibitem{Ramakrishnan:2014}
N.~Ramakrishnan, P.~Butler, S.~Muthiah, and et~al.
\newblock {``Beating the News'' with EMBERS: Forecasting Civil Unrest Using
  Open Source Indicators}.
\newblock In {\em Proceedings of the 20th ACM SIGKDD International Conference
  on Knowledge Discovery and Data Mining}, KDD, pages 1799--1808, New York, NY,
  USA, 2014.

\bibitem{Ritterman09}
J.~Ritterman, M.~Osborne, and E.~Klein.
\newblock Using prediction markets and twitter to predict a swine flu pandemic.
\newblock In {\em Proceedings of the 1st International Workshop on Mining
  Social}, 2009.

\bibitem{Rong:2015:WHI}
Y.~Rong, H.~Cheng, and Z.~Mo.
\newblock Why it happened: Identifying and modeling the reasons of the
  happening of social events.
\newblock In {\em Proceedings of the 21th ACM SIGKDD International Conference
  on Knowledge Discovery and Data Mining}, KDD '15, pages 1015--1024, New York,
  NY, USA, 2015. ACM.

\bibitem{tumasjan2010predicting}
A.~Tumasjan, T.~Sprenger, P.~Sandner, and I.~Welpe.
\newblock Predicting elections with twitter: What 140 characters reveal about
  political sentiment.
\newblock In {\em Proceedings of the Fourth International AAAI Conference on
  Weblogs and Social Media}, pages 178--185, 2010.

\bibitem{Wang:2012}
X.~Wang, M.~S. Gerber, and D.~E. Brown.
\newblock Automatic crime prediction using events extracted from twitter posts.
\newblock In {\em Proceedings of the 5th International Conference on Social
  Computing, Behavioral-Cultural Modeling and Prediction}, SBP, pages 231--238,
  Berlin, Heidelberg, 2012.

\bibitem{WangZYB15}
X.~Wang, Z.~Zhu, C.~Yao, and X.~Bai.
\newblock Relaxed multiple-instance {SVM} with application to object discovery.
\newblock {\em CoRR}, abs/1510.01027, 2015.

\bibitem{Weidmann2003mil}
N.~Weidmann, E.~Frank, and B.~Pfahringer.
\newblock A two-level learning method for generalized multi-instance problems.
\newblock In {\em The European Conference on Machine Learning and Principles
  and Practice of Knowledge Discovery in Databases}, pages 468--479, 2003.

\bibitem{zhao15:sdm}
L.~Zhao, F.~Chen, C.~Lu, and N.~Ramakrishnan.
\newblock Spatiotemporal event forecasting in social media.
\newblock In {\em Proceedings of the {SIAM} International Conference on Data
  Mining, Vancouver, BC, Canada}, pages 963--971, 2015.

\bibitem{Zhao:2015}
L.~Zhao, Q.~Sun, J.~Ye, F.~Chen, C.-T. Lu, and N.~Ramakrishnan.
\newblock Multi-task learning for spatio-temporal event forecasting.
\newblock In {\em Proceedings of the 21th ACM SIGKDD International Conference
  on Knowledge Discovery and Data Mining}, KDD, pages 1503--1512, New York, NY,
  USA, 2015.

\bibitem{zhou07}
Z.-H. Zhou and J.-M. Xu.
\newblock On the relation between multi-instance learning and semi-supervised
  learning.
\newblock In {\em Proceedings of the 24th International Conference on Machine
  Learning (ICML)}, volume 227, pages 1167--1174, 2007.

\end{thebibliography}

\end{document}